\documentclass{article}
\usepackage[utf8]{inputenc}
\usepackage{subfig}
\usepackage{authblk}
\usepackage{amsmath}
\usepackage{amssymb}
\usepackage{graphicx}% Include figure files
\usepackage{dcolumn}% Align table columns on decimal point
\usepackage{bm}% bold math
\usepackage{geometry}
\usepackage{cite}
\title{Net energy up-conversion processes in CdSe/CdS (core/shell) quantum dots, a possible pathway to towards optical cooling}
\author{Muchuan Hua and Ricardo S. Decca\footnote{rdecca@iupui.edu}}
\affil{Department of Physics, Indiana University-Purdue University Indianapolis}
\date{March 2022}

\begin{document}

\maketitle
\begin{abstract}
The investigation of the possibility of optical refrigeration (OR) on zinc-blende cadmium selenide/cadmium sulfide (CdSe/CdS) core/shell structure quantum dots (QDs) has been carried out. Quality samples were synthesized in our lab, and significant energy up-conversion photoluminescence (UCPL) was observed in these samples, showing the potential of generating net cooling effects. To better understand and predict the UCPL characteristics of the QDs, a semi-empirical model has been developed, showing good agreement with our experimental results. The model takes into account the  corresponding quantum yield and cooling efficiency, predicting the possibility of realizing optical refrigeration on a CdSe QDs system.
\end{abstract}
\section{\label{sec:level1}Introduction}
The idea of OR in solids was first proposed by Pringsheim in $1929$.\cite{Pringsheim:1929} He suggested that the thermal vibration energy in solids can be removed by spontaneous anti-Stokes (energy up-conversion) photoluminescence (PL), where the mean emission energy, $\bar{\varepsilon}_{\textrm{em}}$, is higher than the energy of the excitation light, $\varepsilon_{\textrm{ex}}$. However, implementations of this technique are very limited, and most of them are in rare-earth-ion doped glass systems.\cite{epstein:1995,Clark:1996,Mungan:1997,Bowman:2000,Mendioroz:2002,Hoyt:2003,Bahae:2004,Bigotta:2006,Fernandez:2006,Patterson:2008} In 2012, Zhang \textit{et al.}showed the possibility of realizing OR in semiconducting nanomaterials.\cite{Zhang:2012}  In their experiment, CdS nanobelts were cooled by $40$~K under laser light. It is attractive to test OR in other semiconducting nanomaterials, such as semiconducting quantum dots (QDs). Semiconducting QDs are well known for their optical properties due to the ``quantum size effect", {\it i.~e.} quantization of the absorption spectra and size-tunable energy band gap.\cite{EKIMOV:1985} These phenomena allow wide applications of semiconducting QDs in light emitting and photovoltaic devices.\cite{Klimov:2000,Azmi:2016,QLED}
It would be attractive to use the tunable energy gap and ``atomic"-like emission spectra from QDs to achieve OR, particularly when considering their large absorption cross-section. Besides being a natural extension to the work done in CdS nanobelts, QDs would provide a much more versatile platform for OR, since it is simpler to prepare them in suspension or in a solid matrix.\cite{Zhang:2012}

To achieve OR in a material, the cooling efficiency $\eta_{\textrm{c}}$
\begin{equation}
    \eta_{\textrm{c}}=\eta_{\textrm{abs}}\eta_{\textrm{ext}}\frac{\bar{\varepsilon}_{\textrm{em}}}{\varepsilon_{\textrm{ex}}}-1,
    \label{eff}
\end{equation}
must be positive \cite{Bahae:2004}. Here $\eta_{\textrm{abs}}$ is the absorption efficiency of the system, defined as $\eta_{\textrm{abs}}=\frac{\alpha(\varepsilon_{\textrm{ex}},T)}{\alpha(\varepsilon_{\textrm{ex}},T)+\alpha_{\textrm{b}}(\varepsilon_{\textrm{ex}},T)}$, where $\alpha(\varepsilon_{\textrm{ex}},T)$ and $\alpha_{\textrm{b}}(\varepsilon_{\textrm{ex}},T)$ denote the cooling material and system background's light absorption rates respectively. $\eta_{\textrm{ext}}=\eta\eta_{\textrm{es}}$, where $\eta$ denotes the PL quantum yield (QY) of the system, while $\eta_{\textrm{es}}$ is the PL escape efficiency, defined as the likelihood of a emitted photon leaving the system. By considering a fairly optimized system, where both $\eta_{\textrm{abs}}$ and $\eta_{\textrm{ext}}$ approach unity, Eq. \ref{eff} becomes,
\begin{equation}
    \eta_{\textrm{c}}\approx\eta\frac{\bar{\varepsilon}_{\textrm{em}}}{\varepsilon_{\textrm{ex}}}-1.
    \label{eq:ceo}
\end{equation}
Consequently, having a nearly unity QY and net energy up-conversion (UC) during PL processes are critical to realizing OR. However, since QDs' sizes are typically comparable to their lattice size (a few to tens of lattice parameters along each axis), uncoordinated atoms usually exist on QDs' surfaces, which form ion traps (also known as surface traps).\cite{surfacetrap:1986,Califano:2003,surfacetrap:2015,Singh:2018,Zou:2018} These surface traps have long trapping lifetimes, favor non-radiative decay processes, introducing a significant drawback in QDs' QY.\cite{surfacelifetime:2003,lifetime:2006} Therefore, OR was considered unlikely in QDs. Around 2015, multiple breakthroughs were made in CdSe/CdS (core/shell structure) QDs synthesis, where samples with a QY close to unity were produced.\cite{Igor:2016,Peng:2016} According to Refs. \cite{Nan:2012,Igor:2016,Peng:2016}, complete surface passivation on CdSe QDs was achieved by coating them with a CdS shell followed by extra ligand treatment. These reports increase the likelihood of realizing OR in CdSe/CdS QDs.

This work focuses on another important criterion for OR, the capability of generating net energy UC during PL processes in CdSe QDs. Typical PL spectra of QDs are obtained with excitation energy, $\varepsilon_{\textrm{ex}}$ much higher than the QDs' absorption band-edge (referred as HPL spectra), where the radiative recombinations of excitons are mostly through the QDs' intrinsic electronic states (or known as electronic band states).\cite{Brus:1986,Mori:1989,Klein:1990,Efros:1992,Efros:1996,Kelley:2010,Ye:2011, Zhang:2012} In contrast, to realize OR in a solid system, for instant, $\textrm{Yb}^{3+}$-doped glass, $\varepsilon_{\textrm{ex}}$ is expected to be the lowest allowed absorption energy, such that the state density distribution (SDD) is favoring the energy UC processes, as the energy down-conversion (DC) processes are always thermally more favorable. In the case of QDs, their HPL spectra are universally red-shifted from the absorption edge at non-cryogenic temperature, meaning $\varepsilon_{\textrm{ex}}$ needs to be much lower (depends on the materials and size of the QDs, typically, about 80 meV for CdSe/CdS QDs) than the absorption band-edge in order to generate net energy UC.\cite{others:2009, others:2010, others:2013} The existence of such up-conversion photoluminescence (UCPL) processes in group \MakeUppercase{\romannumeral 2}-\MakeUppercase{\romannumeral 6} QDs have been previously observed and the QDs' surface states (Ref. \cite{Wang:2003}) and inter-facial states between the core and the coating shell (Ref. \cite{Rusakov:2003}) were concluded as the origins of the UCPL processes.\cite{Poles:1999,Underwood:2001,Rakovich:2002,Rusakov:2003,Wang:2003} As the whole synthesizing procedure to produce near unity QY colloidal QDs ties to remove the internal and surface defects, such UCPL processes could also be completely prohibited.\cite{Nan:2012,Peng:2016} Thus, characterizing the high quality CdSe/CdS QDs' PL properties with sub-band excitation (SBE) is crucial to OR. In this work, zinc-blende CdSe QDs coated with zinc-blende CdS shells were synthesized. They were reported to have almost unity QY and to be mass production friendly, allowing their wide applications.\cite{Bullen:2004,Chen:2008,Nan:2012} Based on the experimental data collected with these samples, a phenomenological model to describe the UCPL processes in zinc-blende CdSe/CdS QDs is proposed to help estimate the possible cooling efficiency that such QD systems could achieve.
In the rest of the paper, in Section \ref{experimental} details of the experimental techniques for sample characterization and the apparatus used in obtaining PL data are discussed. Special emphasis is placed on the modifications from usual PL systems that permit the collection of the PL signal in the samples. In Section \ref{results} the obtained experimental data is shown. In Section \ref{model} the model for the recombination processes based on the experimental results is presented, where different dimensionality for the vibrational modes are considered. Finally in Section \ref{conclusions} the conclusions of our work are presented.

\section{Experimental setup}\label{experimental}
Experimental data were obtained from CdSe/CdS core/shell structure QDs samples prepared in our laboratory. Synthesis of the samples were carried out in two stages: synthesis of CdSe seeds and growth of the CdS shell. The method for seeds synthesis is well documented,\cite{Karthish} the shell growth was accomplished by using a single precursor (cadmium diethyldithiocarbamate) with an adhesion-growth method,\cite{Nan:2012} and outside the CdS shell, a monolayer of cadmium formate was applied to passivate the shell surface.\cite{Peng:2016, Muchuan:2019}

Absorption spectra of the samples were obtained by using a Thermal Scientific Evolution 600 UV-Vis spectrometer. PL spectra of the QD samples were 
\begin{figure}
    \centering
    \includegraphics[scale=0.5]{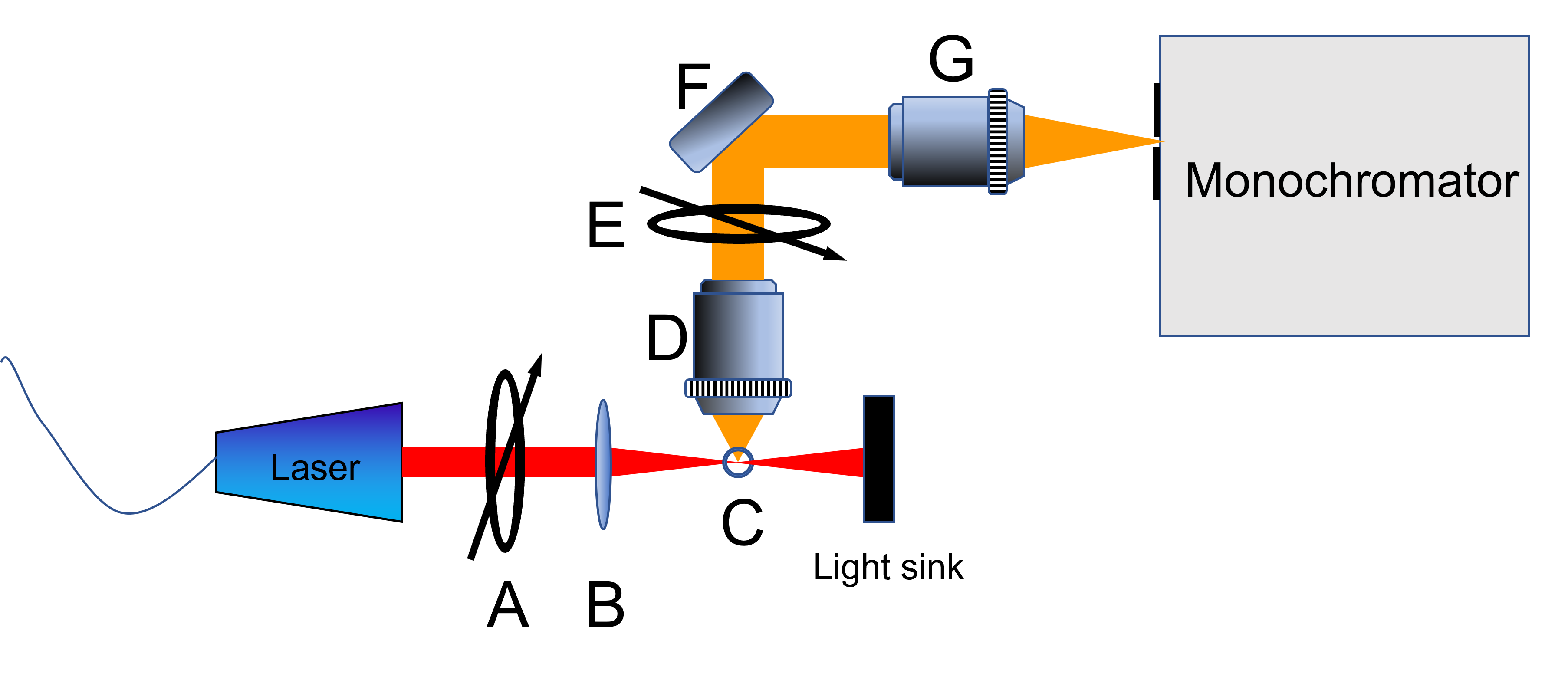}
    \caption{Schematic plot of the off-axis collecting system: A: Polarizer, B: Focusing lens, C: Sample holder, D: Collection objective, E: Analyzer, F: High reflective mirror, G: Adapting objective.}
    \label{ofac}
\end{figure}
obtained by a Horiba Triax 550 monochromator equipped with a $\textrm{LN}_{2}$ cooled CCD chip (CCD 3000). A HeNe laser with emission peak at $1.957$~eV, a laser diode with a tunable emission energy range of $1.919\sim1.945$~eV and another laser diode with a typical emission energy of $3.06\,\textrm{eV}$ were used as the excitation sources.

The common difficulty of observing QD samples' PL spectra at SBE is that the excitation energies are within the range of the PL emission spectra and had an extremely low absorption efficiency, such that the PL signal is overwhelmed by the scattered laser signal. For our application, a full PL line shape evaluation is critical as the possible cooling efficiency is directly related to the spectral profile. In order to increase the PL-to-laser-signal ratio, an off-axis-collecting system was used. As shown in Fig. \ref{ofac}, the PL light of the QDs sample was collected perpendicularly to the incident laser. To further suppress the scattered excitation light, a pair of linear polarizers were introduced.  The mechanism of the system is based on the fact that the PL property of zinc-blende (cubic crystal structure) CdSe QDs has no polarization preference, but the strength of the laser light does (dominated by the Rayleigh scattering). Thus, two polarizers were introduced in the excitation and collection paths separately (part A and E in Fig. \ref{ofac}) to utilize the difference between their polarization properties. One of them was placed right before the focusing lens to horizontally polarize the incident laser light. The other one was used as an analyzer, placed right above the collection objective with its polarization direction aligned with the incident laser beam. Although in such configuration the collecting objective is aligned with the direction of the maximized scattering strength, the polarization of the scattered light is preferentially perpendicular to the analizer. Experimentally, an extinction ratio of about $127:1$  was typically achieved. The usual excitation power was less than $300\,\mu\textrm{W}$ (around $600\,\textrm{W/cm}^2$), such that the mean excitonic density in each QD is much less than one, precluding multi-photon processes.

\section{Experimental data}\label{results}
\begin{figure}
   \centering
   \subfloat[]{%
	    \includegraphics[scale=0.18]{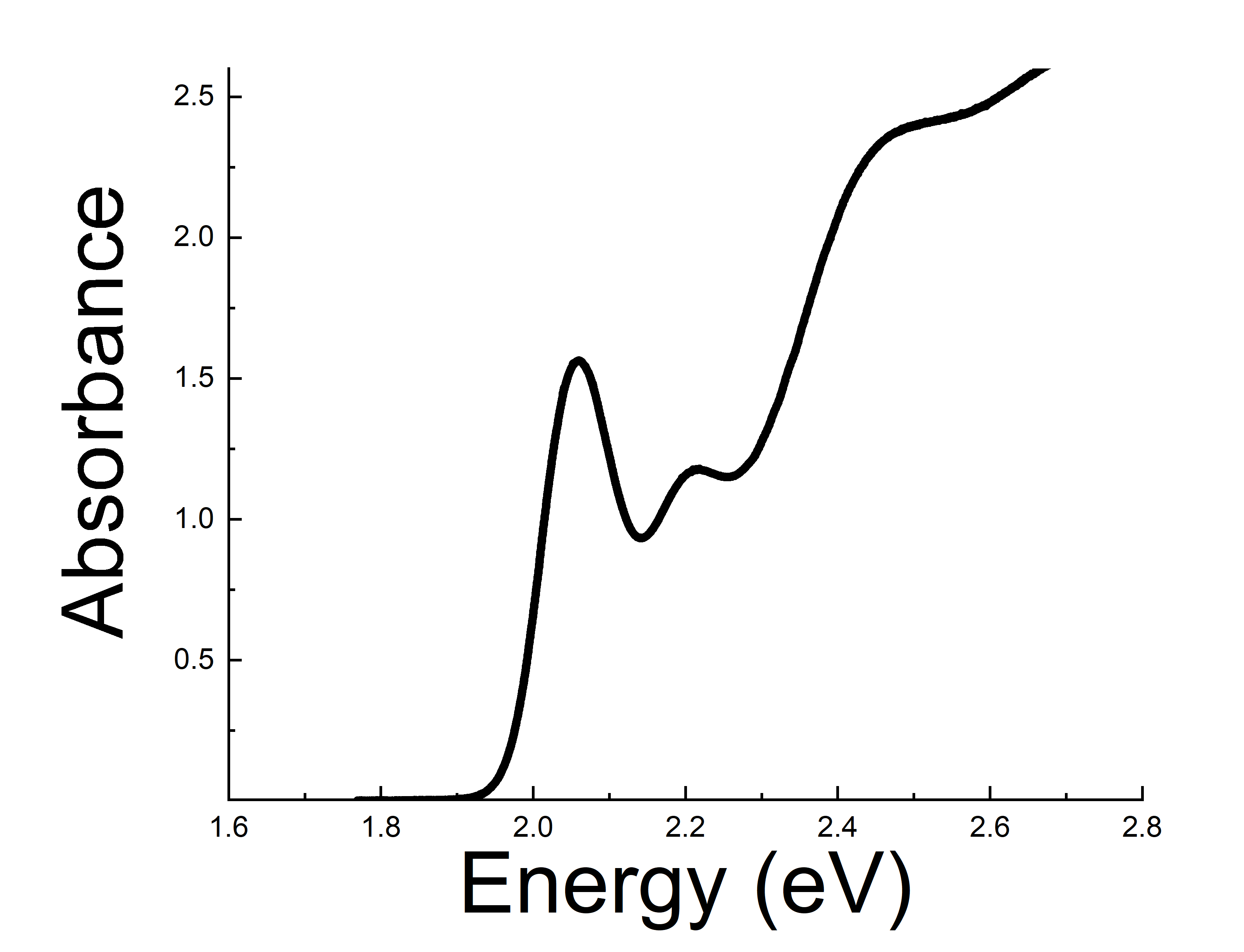}
	    \label{abs}}
	    \quad
        \subfloat[]{%
	    \includegraphics[scale=0.19]{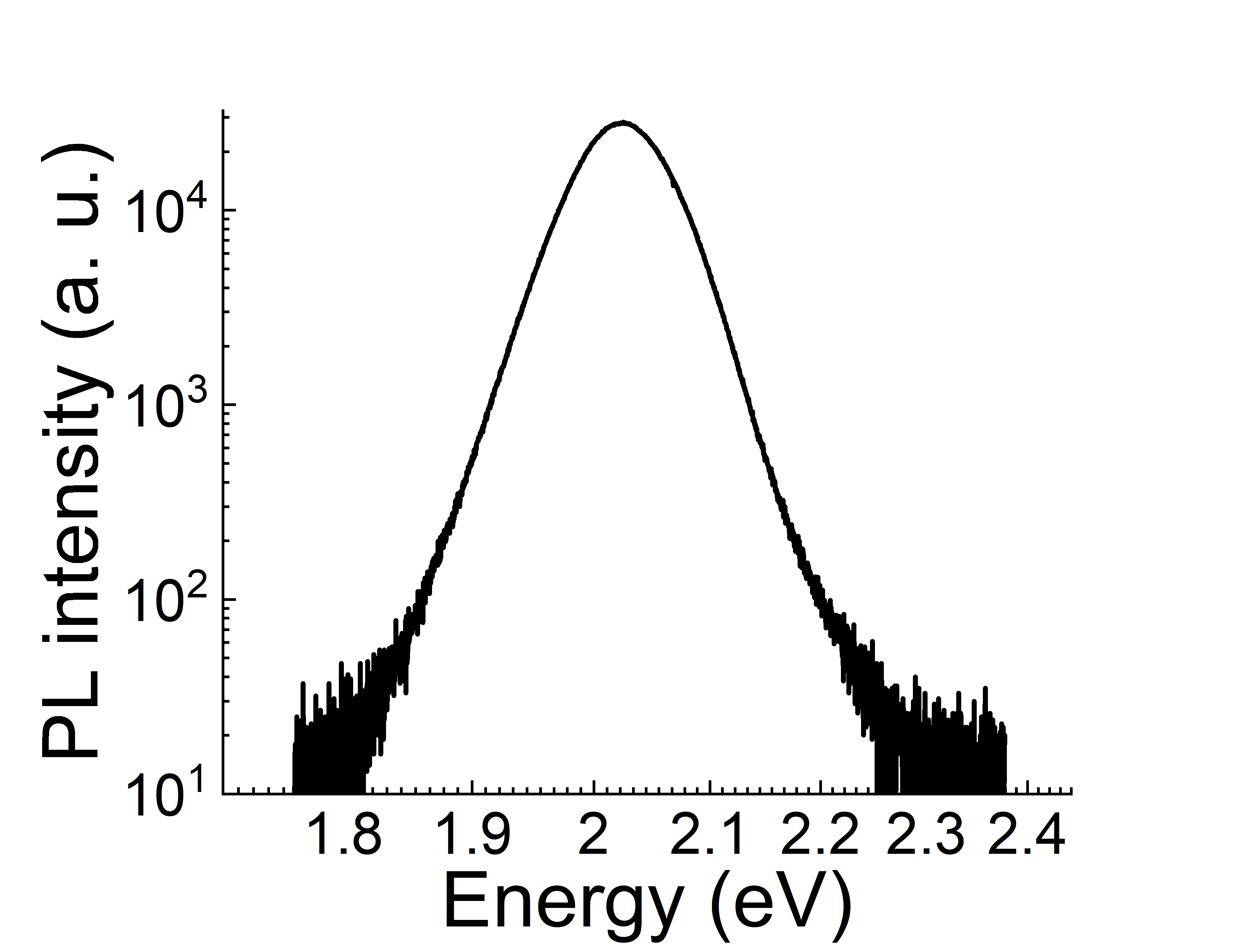}
	    \label{hpl}}
	    \quad
	    \subfloat[]{%
        \includegraphics[scale=0.18]{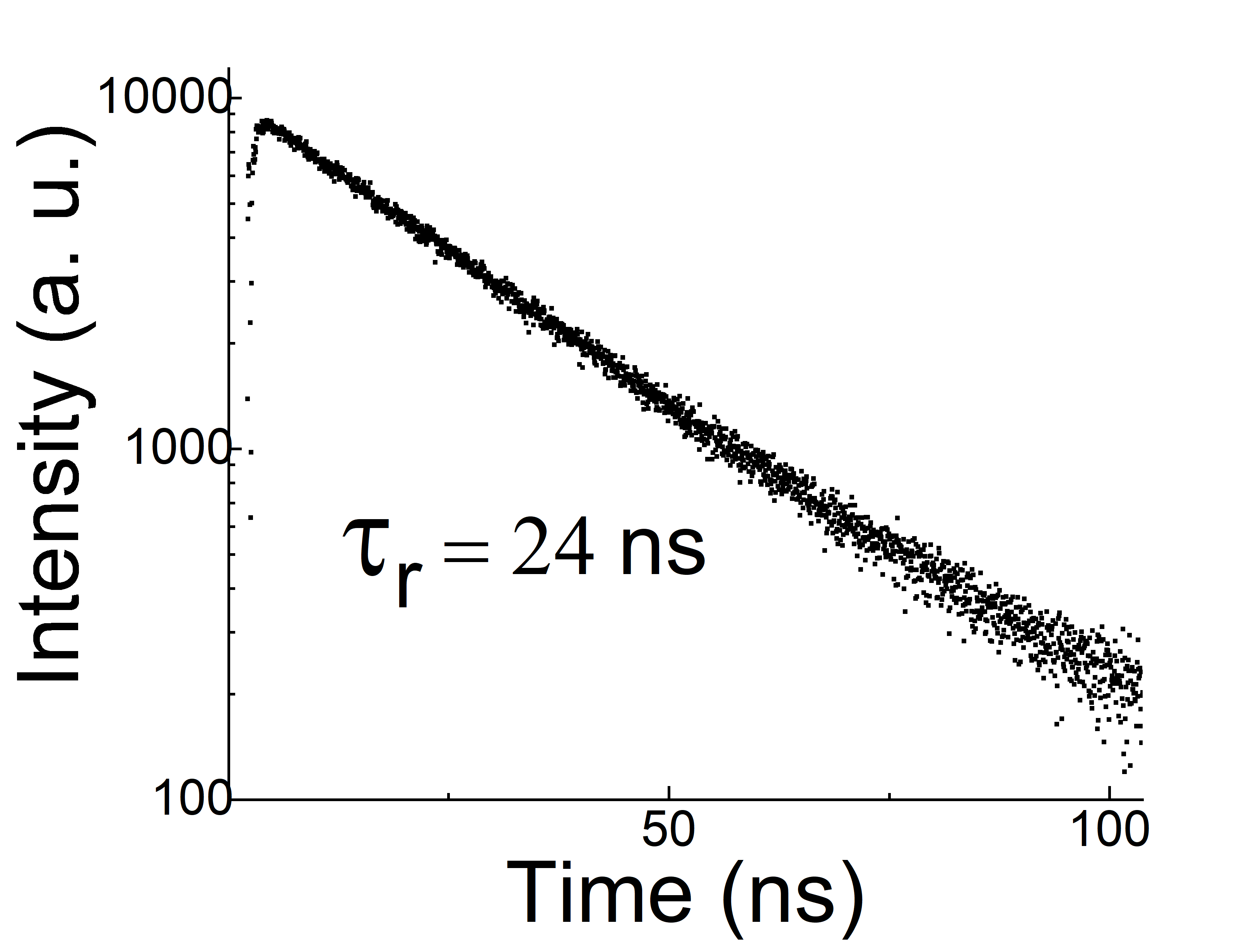}
        \label{dlt}}
        \quad
    \caption{(a) Absorption spectrum of the sample. (b) High energy excitation PL spectrum in logarithmic scale ($\varepsilon_{\textrm{ex}}=3.06\,\textrm{eV}$). (c) Time dependent radiative decay intensity curve of the QDs sample. The curve was fitted with a single exponential curve with a decay lifetime $\tau=24$~ns. All the data shown were collected with sample 3.}
\end{figure}

The basic optical properties of the samples are listed in table \ref{qdsi}.
\begin{table}[h]
    \centering
    \begin{tabular}{c|c|c|c|c}
    \hline
    Sample  & shell thickness & First absorption maximum & HPL peak energy & FWHM  \\
    &  (monolayers) & (eV) & (eV) & (meV) \\
         \hline
         1 & 2 &  & 2.067 & 89\\
         
         2 & 4 &  & 2.027 & 89\\
         
         3 & 4 + CdFt$^{1}$ &  2.059 & 2.023 & 87\\
         4 & 4 + CdFt & 2.049 & 2.011 & 81\\
         
    \end{tabular}
    \caption{PL information of the QDs samples. 1). Surface of the CdS is passivated with a monolayer of CdFt. }
    \label{qdsi}
\end{table}
Sample 1 to 3 were produced from the same batch, while sample 4 was obtained from another batch. Before growing the shells, core sizes were calculated using a semi-empirical equation from Ref. \cite{Karel:2010}, where a value of $3.0\,\textrm{nm}$ were derived for sample 1 to 3, and a value of $3.3\,\textrm{nm}$ for sample 4. The samples' absorption spectra showed clear features of energy quantization (a typical absorption spectrum is shown in Fig. \ref{abs}), indicating the successful synthesis of QDs. The PL FWHM is less than $90$~meV, while their HPL spectra shows rapid decay in both high and low energy sides (Fig. \ref{hpl}), indicating a good monodispersity of QDs' size distribution.\cite{zb:2005,zbseo2:2008} Furthermore, the absence of the deep trap emission around $1.8\,\textrm{eV}$ indicating nearly complete passivation of the core surface.\cite{Fengler:2013} Fig. \ref{dlt} shows the radiative decay lifetime yielding a single exponential curve, confirming the report from Peng's group,\cite{Nan:2012} where unity QY was achieved in their QD samples after complete surface passivation and a single radiative decay lifetime was observed. Hence, these QDs are suitable candidates for investigating specific transitions involved in UCPL. 

\begin{figure}[h!]
\subfloat[]{%
	    \includegraphics[scale=0.26]{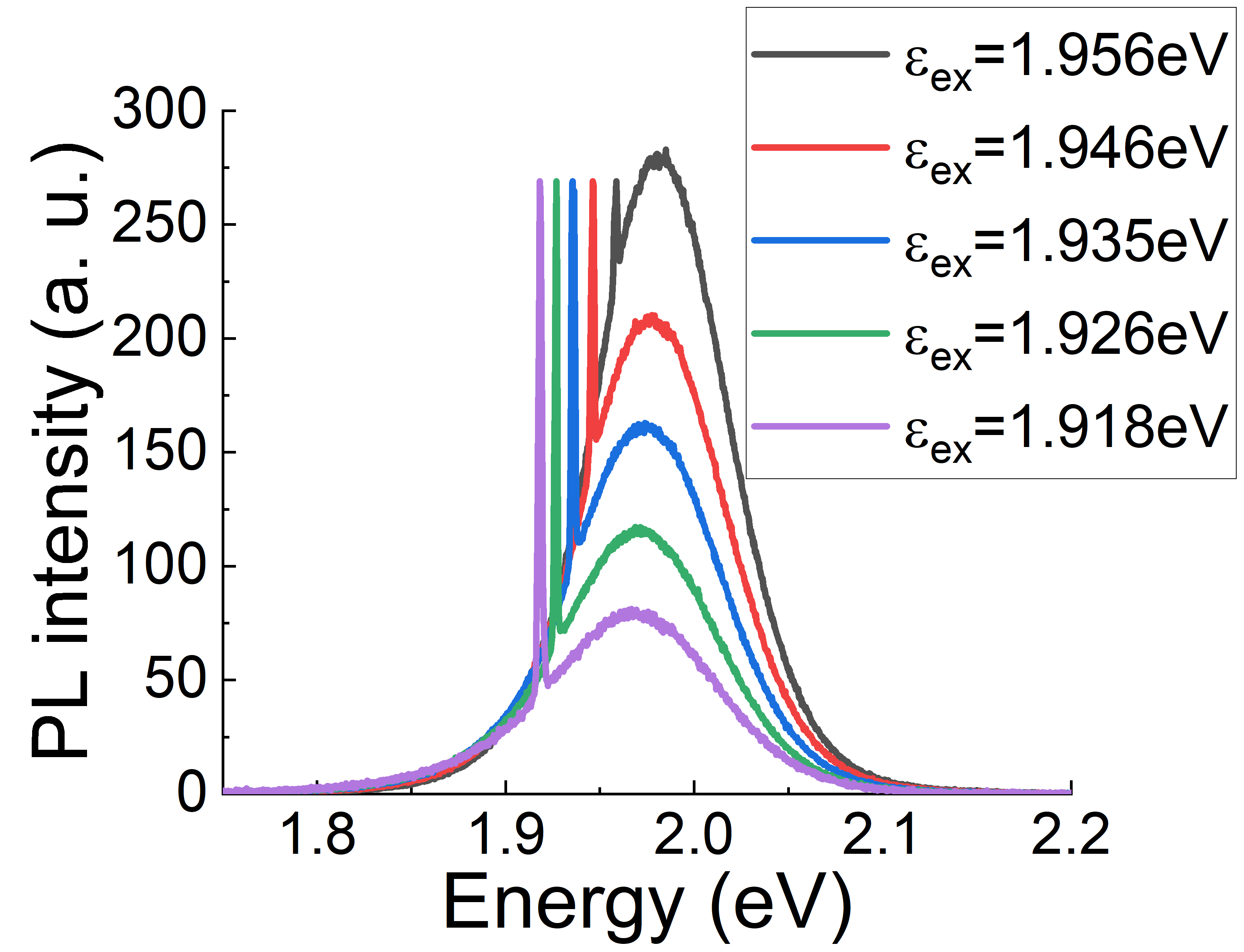}
	    \label{exed}}
	    \quad
	    \subfloat[]{%
        \includegraphics[scale=0.26]{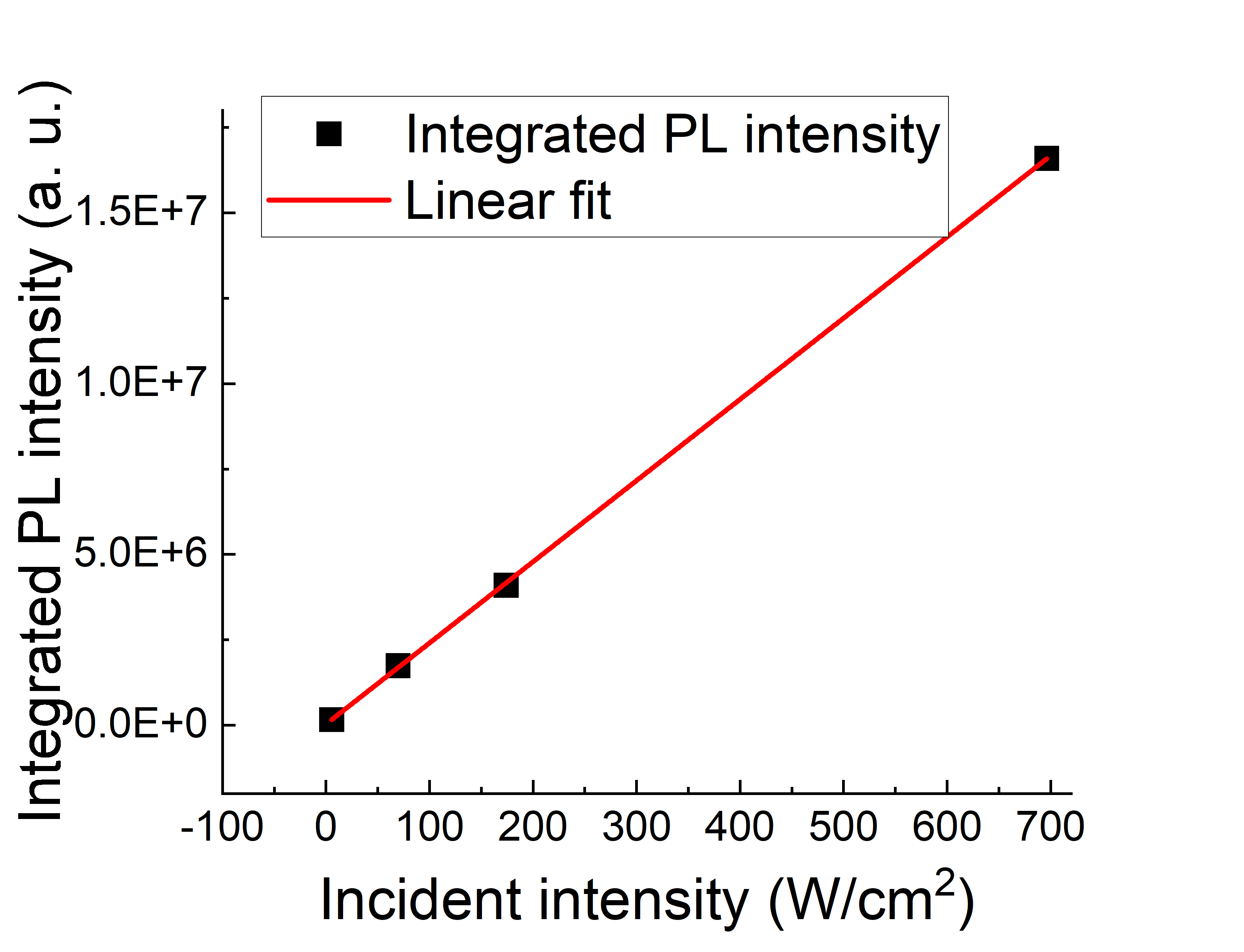}
        \label{ucplid}}%
        \quad
        \caption{(a) Excitation energy dependent PL spectra (PL intensities were normalized with respect to the observed laser signals) of sample 3. (b) Excitation intensity dependence of the PL intensity of sample 4 at $\varepsilon_{\textrm{ex}}=1.941\,\textrm{eV}$. The line is a linear fit to the data.}
\end{figure}

With the help of the off-axis collection system, the laser signal within the spectrum is significantly reduced. As seen in Fig. \ref{exed}, only a small ``spike" around the laser energy was observed in the much larger PL signal. Furthermore, the spike was found to be more noticeable at lower excitation energy, where absorption efficiency is lower, indicating it is not originated from PL processes. Therefore, we conclude such spikes were originated from the scattered laser light, and the samples' PL spectra with SBE have a smooth line shape with a single peak. As shown in Fig. \ref{ucplid}, the sample's PL intensity is linearly dependent on the excitation intensity, indicating as aforementioned, that multi-photon absorption processes have a negligible contribution.
More important for our work a significant UCPL signal was observed in our sample, a prerequisite for OR, see Fig. \ref{exed}

Such UCPL has been observed by Rusakov \textit{et al.} \cite{Rusakov:2003} and Wang \textit{et al.} \cite{Wang:2003} within CdSe/ZnS and CdTe colloidal QDs respectively. Each group has proposed a mechanism, which is different from the other one, to explain the observed UCPL processes. 

\begin{figure}
        \subfloat[]{%
	    \includegraphics[scale=0.26]{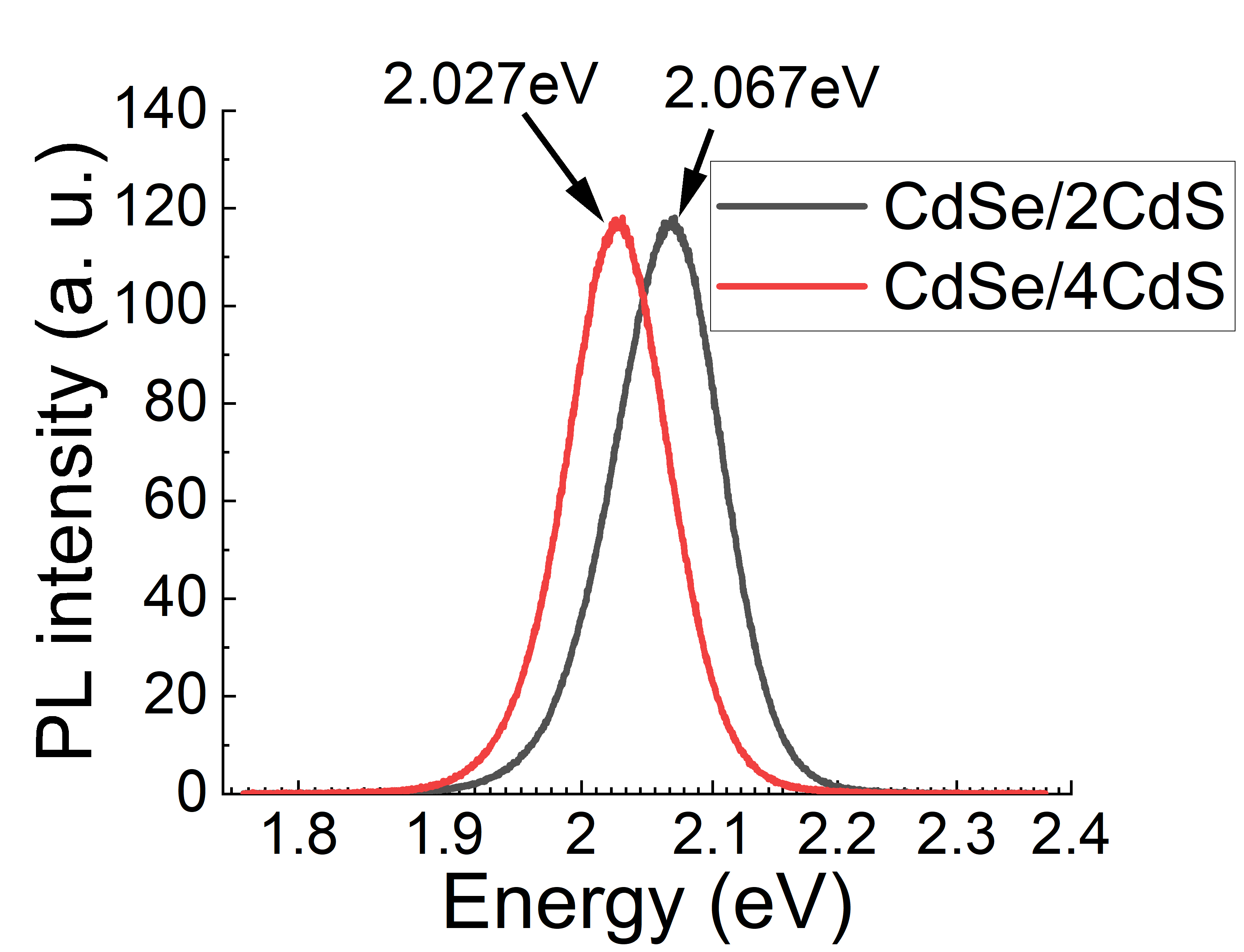}
	    \label{fig:hplshd}}
	    \quad
	    \subfloat[]{%
        \includegraphics[scale=0.26]{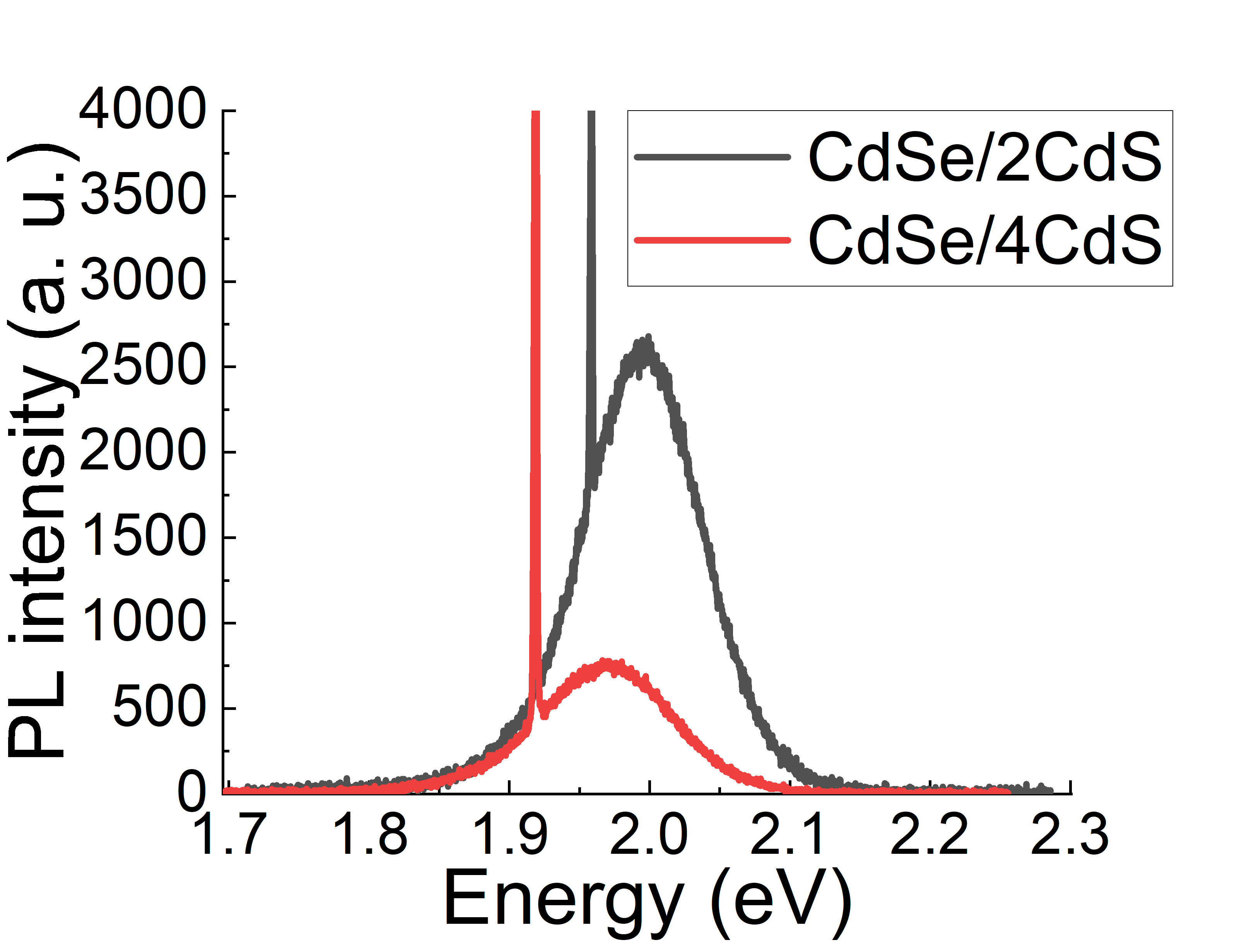}
        \label{fig:ucplstd}}%
        \quad
        \caption{(a) HPL spectra of CdSe/2CdS (black) and CdSe/4CdS (red) samples. The intensities are adjusted to match each other for easy comparison, both samples were not finished with CdFt. (b) PL spectra of CdSe/2CdS (black) and CdSe/4CdS (red) samples excited at $1.956\,\textrm{eV}$ ($91\,\textrm{meV}$ lower than the HPL peak energy) and $1.918\,\textrm{eV}$ ($89\,\textrm{meV}$ lower than its HPL peak energy) respectively. The PL intensities were normalized by their excitation intensities.}
\end{figure}

In Ref. \cite{Rusakov:2003}, the UCPL processes in CdSe/ZnS QDS were concluded to be originated from the trapping states introduced by the inter-facial lattice disorder as the UCPL intensity was enhanced with thicker ZnS shell. 
However, several issues were found when applying this model to our samples: First, lattice mismatch between zinc-blende CdS and zinc-blende CdSe ($3.6\%$) is much smaller than the case of zinc-blende ZnS and zinc-blende CdSe ($10.6\%$), leading to a much lower chance of introducing inter-facial disorder. Second, CdSe/CdS QDs' intrinsic energy gap is supposed to have a strong shell thickness dependence. 
To address the shell thickness issue, we compare the PL spectra of sample $2$ (sample with 4 monolayers thick CdS shell) with sample $1$ ( with 2 monolayers shell). 
As shown in Fig \ref{fig:hplshd}, however, the HPL peak energy has been shifted from $2.067\,\textrm{eV}$ to $2.027\,\textrm{eV}$ after increasing the CdS shell thickness from two to four monolayers. 
Therefore, a potential enhancement observed with a fixed excitation energy could be a consequence of the increased band-edge absorption due to the thicker shell. 
To minimize the effect due to the shifting in the QDs' intrinsic energy band-gap, the comparison was done between the two samples' PL spectra excited with laser energies lowered by an equal amount with respect to their own HPL peak energies. 
As shown in Fig. \ref{fig:ucplstd}, with excitation energies around $90\,\textrm{meV}$ lower than their HPL peak energies, in both samples the UCPL spectra peaked about $40\,\textrm{meV}$ above the excitation energies, suggesting UCPL's size dependence is identical to the HPL. At the same time, the PL intensity dropped significantly as the shell thickness increases, while an increase in the inter-facial disorder is expected. Such observations negate the hypothesis that the UCPL originates from the inter-facial lattice disorder in CdSe/CdS QDs.

\begin{figure}
    \centering
    \includegraphics[scale=0.4]{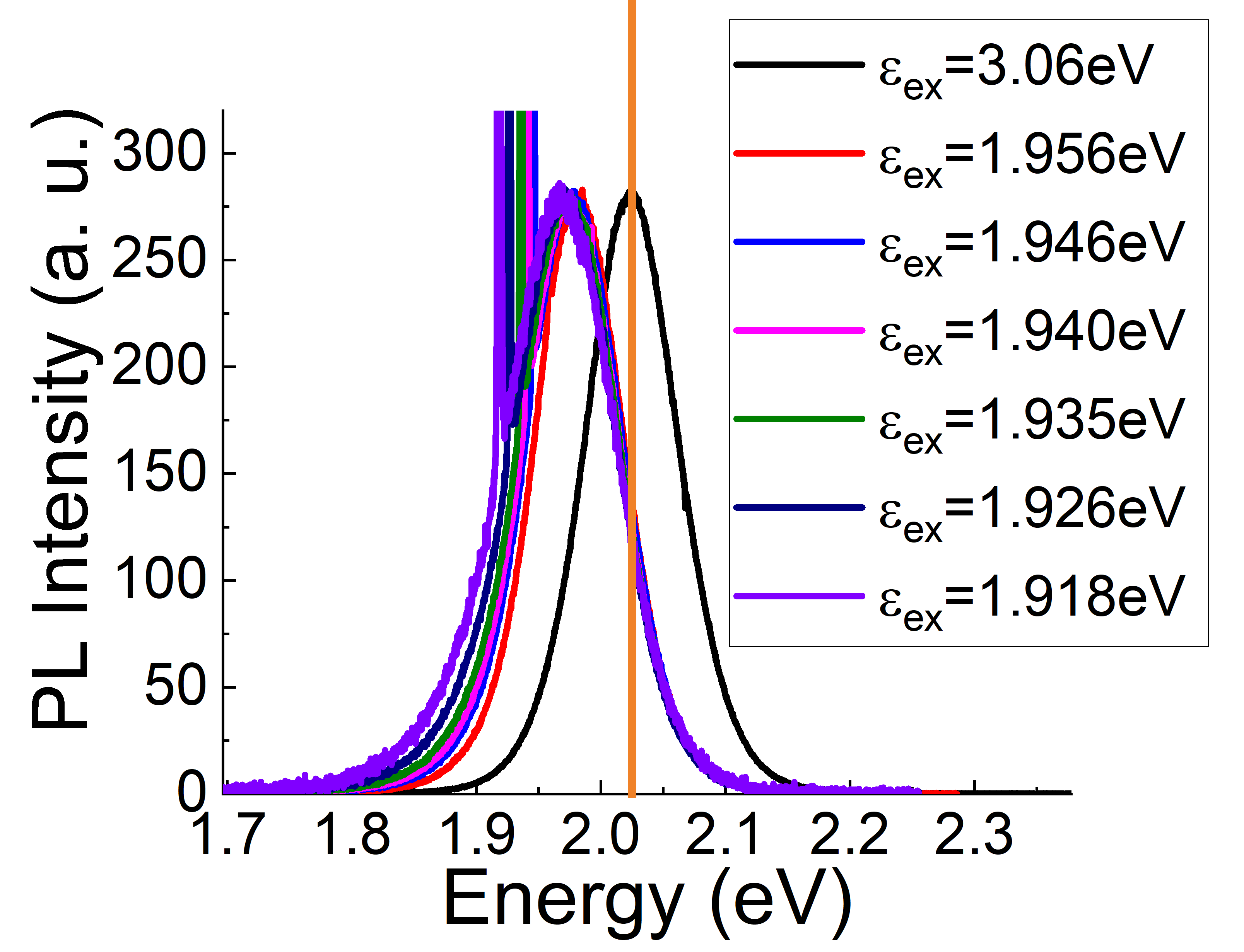}
    \caption{PL spectra of sample 3 with excitation energies at $3.05\,\textrm{eV}$ (black), $1.956\,\textrm{eV}$ (red), $1.946\,\textrm{eV}$ (blue), $1.940\,\textrm{eV}$ (magenta), $1.935\,\textrm{eV}$ (olive), $1.926\,\textrm{eV}$ (navy) and $1.918\,\textrm{eV}$ (purple). For ease of comparison spectral intensities were adjusted to have the same maximum value. The energy of the HPL maximum is marked by the orange line.}
    \label{fig:ucpltp}
\end{figure}

The model proposed in Ref. \cite{Wang:2003}, based on the observation of UCPL in CdTe colloidal QDs, suggests UCPL was generated by photon absorption between shallow surface electron and hole trapping states. In the model, the emission processes involve the thermally excited electron (hole) in the conduction band (valence band)
radiatively recombines with the hole (electron) in the surface trapping states. According to this model, the following PL properties should be expected: 
First, UCPL intensity should decrease as the shell thickness increases, since the QDs' surfaces are spatially farther away from their cores, leading to less overlap between the intrinsic band states and the surface trapping states. 
Second, size dependence of UCPL should be much less pronounced than HPL, as surface states are typically less size sensitive than QDs' intrinsic band gap. 
Third, the possibility of having both electron and hole recombined radiatively from the QDs intrinsic states is determined by the thermal population of both carriers. 
When applied to our sample, by assuming the electron and hole trapping states are equally spaced, about $50\,\textrm{meV}$ from the QDs' conduction and valence band respectively, the chance of having both carriers thermally activated to the intrinsic band states is about $1/10$ (at $293\,\textrm{K}$) of the chance of having only one of them being populated. In our data we see the first expectation is confirmed (Fig. \ref{fig:ucplstd}), while the second (Fig. \ref{fig:hplshd}) and third ones (Fig. \ref{fig:ucpltp}) do not match our results.
Regarding the last two points, the UCPL of our samples showed identical size dependence as their HPL and, as shown in Fig. \ref{fig:ucpltp}, the PL intensities at the HPL peak are around $48.5\%$ of the its maximum regardless of the excitation energy. 
In conclusion, the UCPL observed in our CdSe/CdS QD samples cannot be well described by the existing models used in colloidal QDs. Thus, a new model is proposed based on our experimental data to explain the UCPL processes inside CdSe/CdS QDs.

\section{Model for the PL processes with sub-band-excitation}\label{model}
\begin{figure}
    \centering
    \includegraphics[scale=0.4]{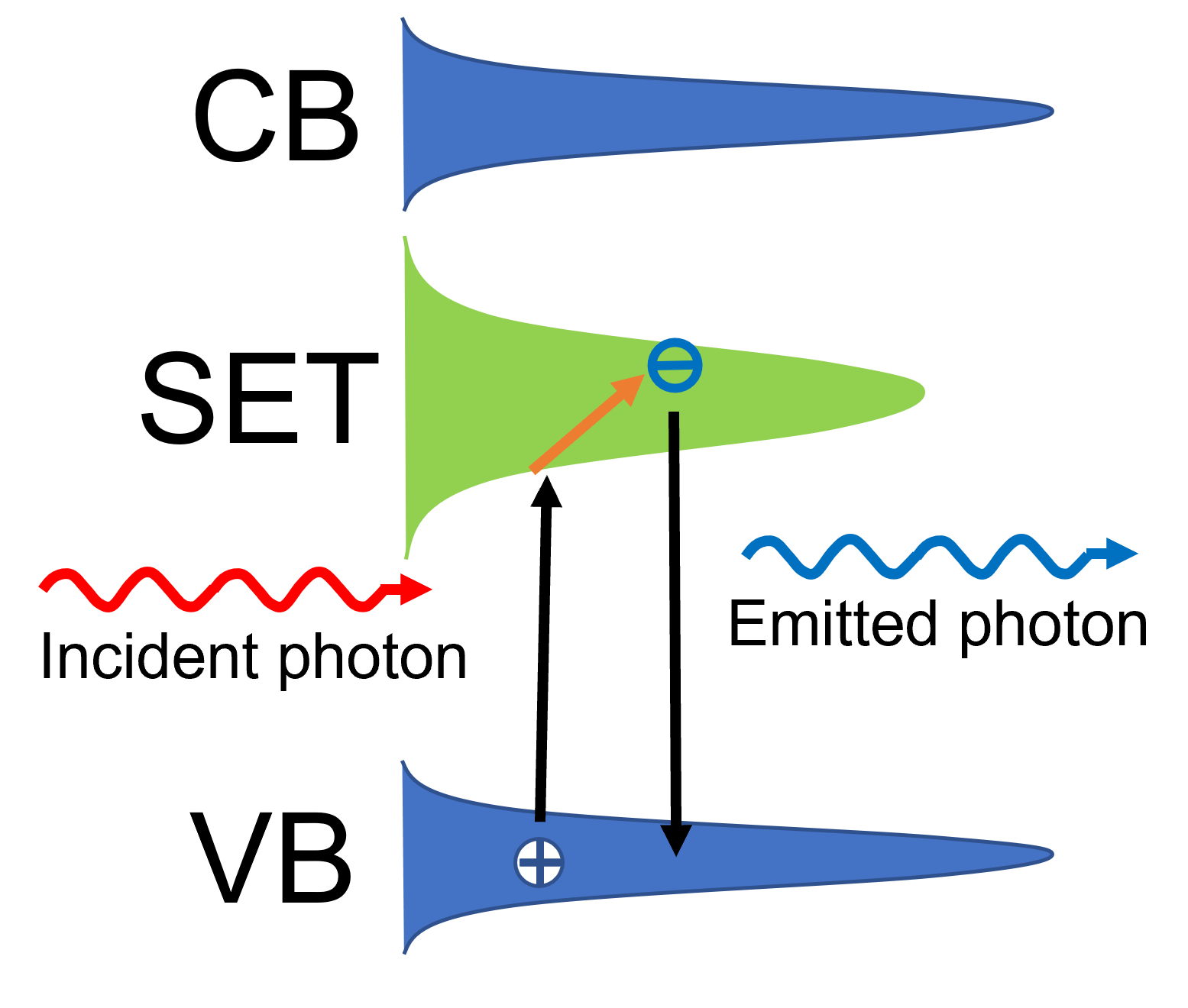}
    \caption{Scheme of excitonic processes involved in the UCPL in CdSe/CdS QDs. CB: bottom of the QDs' intrinsic conduction band. VB: top of the QDs' intrinsic valence band. SET: shallow surface electron trapping states.}
    \label{fig:model}
\end{figure}

One of the key points in our proposed model is associated with the lack of the thermalization of electrons between conduction and trap states. Furthermore,
the main mechanism of carrier thermalization within a band is by acoustic phonon coupling, since longitudinal optical phonon coupling strength in CdSe QDs is much weaker than in bulk.\cite{Alivisatos:1989,LOP:1990,KELLEY:2013}
A possible candidate for energy states responsible for radiative recombination is the surface electron trapping states (SET) of CdS, which lies within $0.2\,\textrm{eV}$ below CdS's intrinsic energy gap.\cite{Jules:1966} The shallow surface hole traps were not taken into account, as the main target of the shell growth and the final surface passivation with the CdFt monolayer is to remove surface hole traps, the major source of non-radiative decay processes.\cite{Aisea:2015,Peng:2016} Hence, the proposed mechanism is shown in Fig \ref{fig:model}. Under SBE, in CdSe/CdS QDs, the photon absorption is achieved by an optical transition between the SET and the QDs' intrinsic valence band edge (VBE). Due to the poor overlap between the SET and VBE, this exciton has a very long lifetime (tens of ns \cite{Wang:2003}), allowing the electron to be thermalized inside the SET. Finally the electron recombine with the hole left in the VBE through radiative or non-radiative decay processes. Following such mechanism, the observed strong size dependence in the UCPL emission energy, and the unusually high UCPL intensity observed in our samples can be properly explained.
Based on this model, the line shape of both UCPL and down-conversion photoluminescence (DCPL) are determined by the thermalization processes inside the SET. In a more realistic case, the broad line shape of QDs are produced from their size distribution, phonon-coupled PL processes, existence of other sub-band defect states and other unknown processes. Therefore, mathematically, a Gaussian distribution, $O_{\textrm{s}}(\varepsilon)$ with center $\varepsilon_{\textrm{s}}$ and variance $\sigma_{\textrm{s}}$ were assigned as the joint excitonic SDD for SET. They were determined by calculating average value of the UCPL peak energies and FWHMs of the experimental data respectively. To complete the overall line shape, three excitonic transition probabilities need to be defined:\\
$k_{\textrm{s}}$: Intra-band transition probability between the states inside the SET;\\
$k_{\textrm{r}}$: Radiative decay probability;\\
$k_{\textrm{nr}}$: Non-radiative decay probability.\\
For simplicity, all transition probabilities were considered to be energy independent. The mechanism for intra-band transitions is attributed to be an acoustic phonon-assisted carrier transfer, where the transition rate is also modified by the phonons' SDD. For the sake of simplicity, the acoustic phonon dispersion relationship used is $$\omega=ck,$$ where $\omega$ is the phonons' angular frequency, $k$ denotes their wave number and $c$ denotes the speed of sound in the material. Although such approximation does not provide a comprehensive description, it is sufficient to evaluate the physics behind the process. Therefore, in an n-dimensional system, the phonon's SDD is
\begin{equation}
    g_{\textrm{n}}(\varepsilon)=C_{\textrm{g}}\left\lvert\frac{\varepsilon^{\textrm{n}-1}}{e^{\varepsilon/k_{\textrm{B}}T}-1}\right\rvert,
    \label{eq:phc}
\end{equation}
where $k_{\textrm{B}}$ is Boltzmann's constant and $C_{\textrm{g}}$ is the coupling amplitude, which contains information of $c$ and was set to one, as it can be treated as a part of $k_{\textrm{s}}$. Thus the time evolution of the excitonic density in the SET, $N_{\textrm{s}}(\varepsilon)$ is
\begin{multline}
    \frac{dN_{\textrm{s}}(\varepsilon)}{dt}=\beta\delta(\varepsilon-\varepsilon_{\textrm{ex}})-k_{\textrm{s}}\int_{\varepsilon-\varepsilon_{c}}^{\varepsilon+\varepsilon_{c}}g_{\textrm{n}}(\varepsilon'-\varepsilon)O_{\textrm{s}}(\varepsilon')d\varepsilon'N_{\textrm{s}}(\varepsilon)\\+k_{\textrm{s}}\int_{\varepsilon-\varepsilon_{c}}^{\varepsilon+\varepsilon_{c}}g_{\textrm{n}}(\varepsilon-\varepsilon')N_{\textrm{s}}(\varepsilon')d\varepsilon'O_{\textrm{s}}(\varepsilon)-(k_{\textrm{r}}+k_{\textrm{nr}})N_{\textrm{s}}(\varepsilon).
    \label{eq:std}
\end{multline}
The terms $\beta\delta(\varepsilon-\varepsilon_{ex})$ and $(k_{\textrm{r}}+k_{\textrm{nr}})N_{\textrm{s}}(\varepsilon)$ describe the photon absorption and emission processes respectively, where $\beta$ is the photon absorption rate and $\delta$ denotes the Dirac delta function. The second and third terms on the right side of Eq. \ref{eq:std} represent $N_{\textrm{s}}(\varepsilon)$ losing and receiving excitons through intra-band transitions respectively. Here $\varepsilon_{c}$ denotes the phonon cutoff energy, not known for QDs at room temperature as the anharmonicity is significant, greatly expanding the phonon spectrum. Experimentally, a broad PL line shape was observed even at SBE. A $\varepsilon_{\textrm{c}}$ value larger than $100$~meV is required to describe the observed broadening, which is significantly larger than the phonon cutoff energy of the bulk material and is not currently well justified. Since $100$~meV is significantly larger than $k_{\textrm{B}}T$, $g(\varepsilon)$ vanishes rapidly when $\varepsilon_{\textrm{c}}$ is larger than the $\textrm{FWHM}$. To simplify  the calculation, $\varepsilon_{\textrm{c}}\rightarrow\infty$ was used in the model. Under such assumption, 
\begin{equation}
    k_{\textrm{s}}\int g_{\textrm{n}}(\varepsilon-\varepsilon')N_{\textrm{s}}(\varepsilon')d\varepsilon'\equiv g_{n}\circledast N_{\textrm{s}}(\varepsilon),
\end{equation}
the convolution between $g_{\textrm{n}}(\varepsilon)$ and $N_{\textrm{s}}(\varepsilon)$. By setting $\int g_{n}(\varepsilon'-\varepsilon)O_{\textrm{s}}(\varepsilon)d\varepsilon'\equiv F(\varepsilon)$, in steady state Eq. \ref{eq:std} becomes
\begin{equation}
    0=k_{\textrm{s}}O_{\textrm{s}}(\varepsilon)g_{n}(\varepsilon)\circledast N_{\textrm{s}}(\varepsilon)-[k_{\textrm{s}}F(\varepsilon)+k_{\textrm{r}}+k_{\textrm{nr}}]N_{\textrm{s}}(\varepsilon)+\beta \delta(\varepsilon-\varepsilon_{\textrm{ex}}).
    \label{nds1}
\end{equation}

In order to derive the expression of $N_{\textrm{s}}(\varepsilon)$, a few approximations were introduced to the calculation. When at most one exciton is present in the QD ($\beta<< k_{\textrm{r}}$), the term $k_{\textrm{s}}O_{\textrm{s}}(\varepsilon)g_{\textrm{n}}(\varepsilon)\circledast N_{\textrm{s}}(\varepsilon)$ is always smaller than the term $k_{\textrm{s}}F(\varepsilon)$. Thus, $k_{\textrm{s}}O_{\textrm{s}}(\varepsilon)g_{\textrm{n}}(\varepsilon)\circledast N_{\textrm{s}}(\varepsilon)$ can be considered as a correction to Eq. \ref{nds1}, and
\begin{equation}
    N_{\textrm{s}}^{0}(\varepsilon)\approx\frac{\beta\delta(\varepsilon-\varepsilon_{\textrm{ex}})}{k_{\textrm{s}}F(\varepsilon)+k_{\textrm{r}}+k_{\textrm{nr}}}.
    \label{eq:ns1}
\end{equation}
The superscript``$0$" denotes the un-modified solution to Eq. \ref{nds1}.
Then $N_{\textrm{s}}(\varepsilon)$ can be solved iteratively by substituting Eq. \ref{eq:ns1} into Eq. \ref{nds1},
\begin{equation}
    0=k_{\textrm{s}}O_{\textrm{s}}(\varepsilon)g_{\textrm{n}}(\varepsilon)\circledast N_{\textrm{s}}^{0}(\varepsilon)-[k_{\textrm{s}}F(\varepsilon)+k_{\textrm{r}}+k_{\textrm{nr}}]N_{\textrm{s}}^{1}(\varepsilon)+\beta\delta(\varepsilon-\varepsilon_{\textrm{ex}}).
\end{equation}
In general, for any integer $\textrm{m}>0$, $N_{\textrm{s}}^{\textrm{m}}(\varepsilon)$ satisfies
\begin{equation}
    0=k_{\textrm{s}}O_{\textrm{s}}(\varepsilon)g_{\textrm{n}}(\varepsilon)\circledast N_{\textrm{s}}^{m-1}(\varepsilon)-[k_{\textrm{s}}F(\varepsilon)+k_{\textrm{r}}+k_{\textrm{nr}}]N_{\textrm{s}}^{m}(\varepsilon)+\beta\delta(\varepsilon-\varepsilon_{\textrm{ex}}).
\end{equation}
The attempt of fitting n started by treating the phonon modes inside the QDs to be identical to the ones in bulk materials, a three dimensional system. Under this assumption, 
\begin{equation}
  g_{3}(\varepsilon)=C_{\textrm{g}}\left\lvert\frac{\varepsilon^{2}}{e^{\varepsilon/k_{\textrm{B}}T}-1}\right\rvert,  
  \label{eq:g3d}
\end{equation}
was plugged into the model. 
\begin{figure}
    \centering
    \subfloat[]{%
        \includegraphics[scale=0.185]{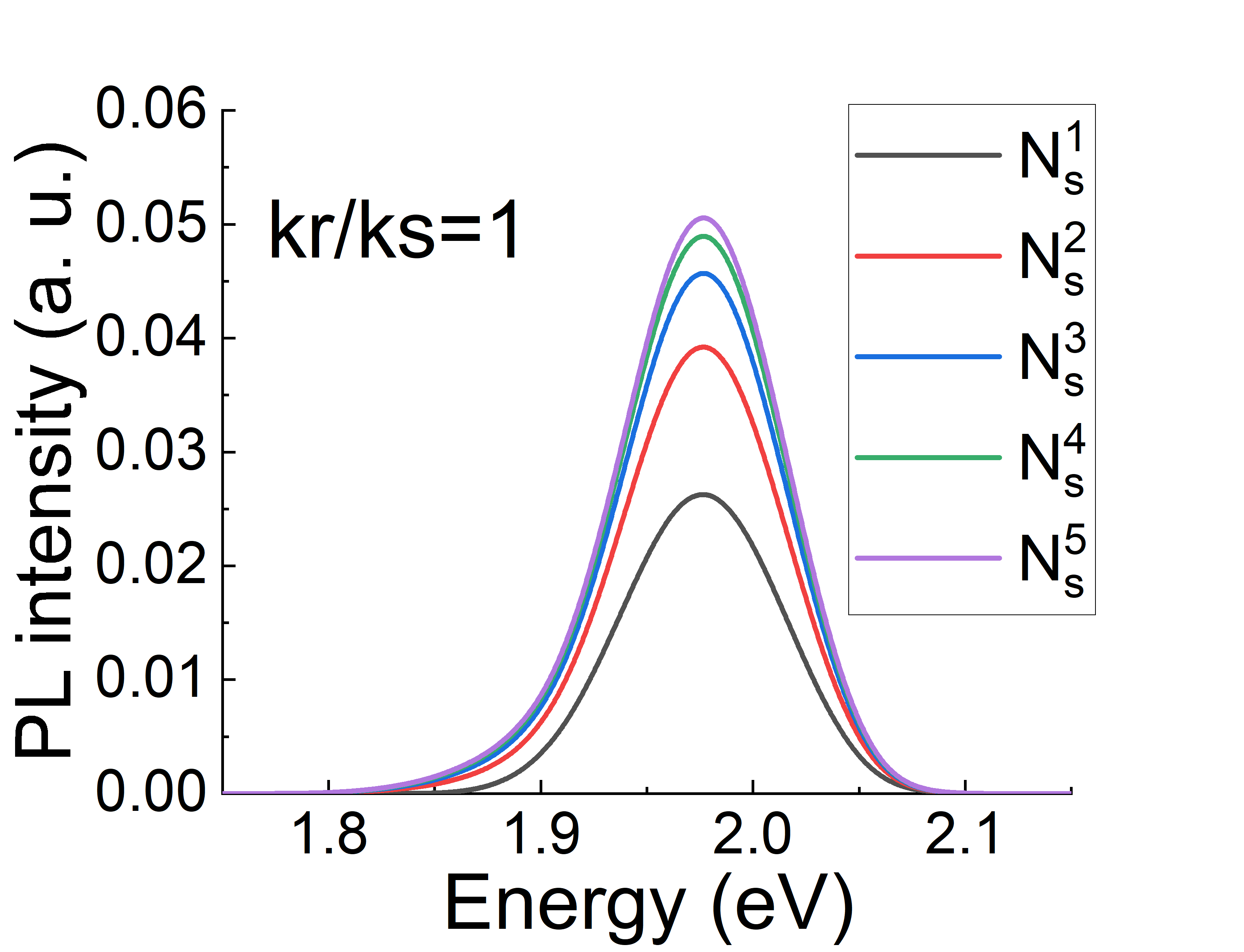}
        \label{fig:ks1}}%
        \quad
    \subfloat[]{%
	    \includegraphics[scale=0.185]{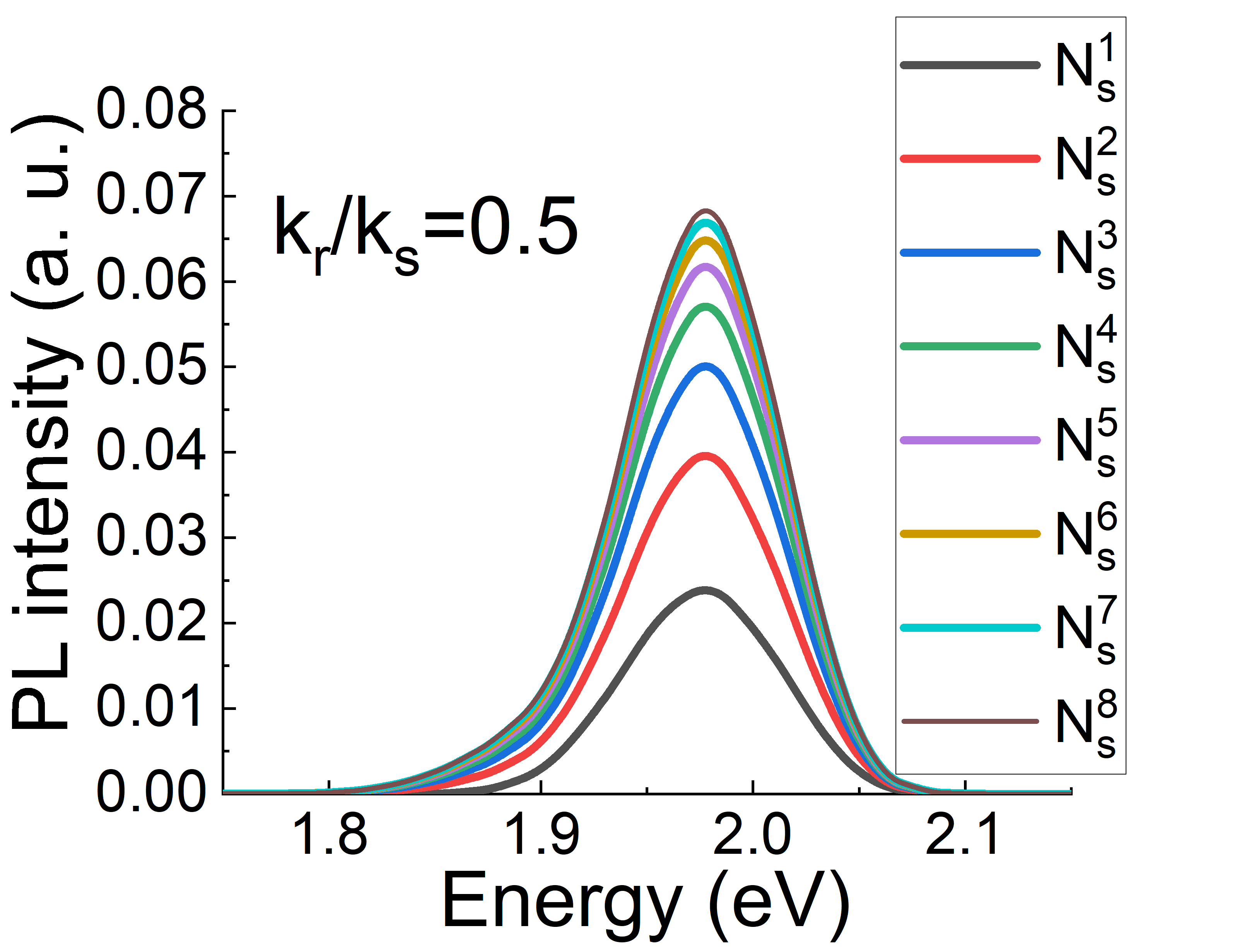}
	    \label{fig:ks05}}
	    \quad
	\subfloat[]{%
	    \includegraphics[scale=0.185]{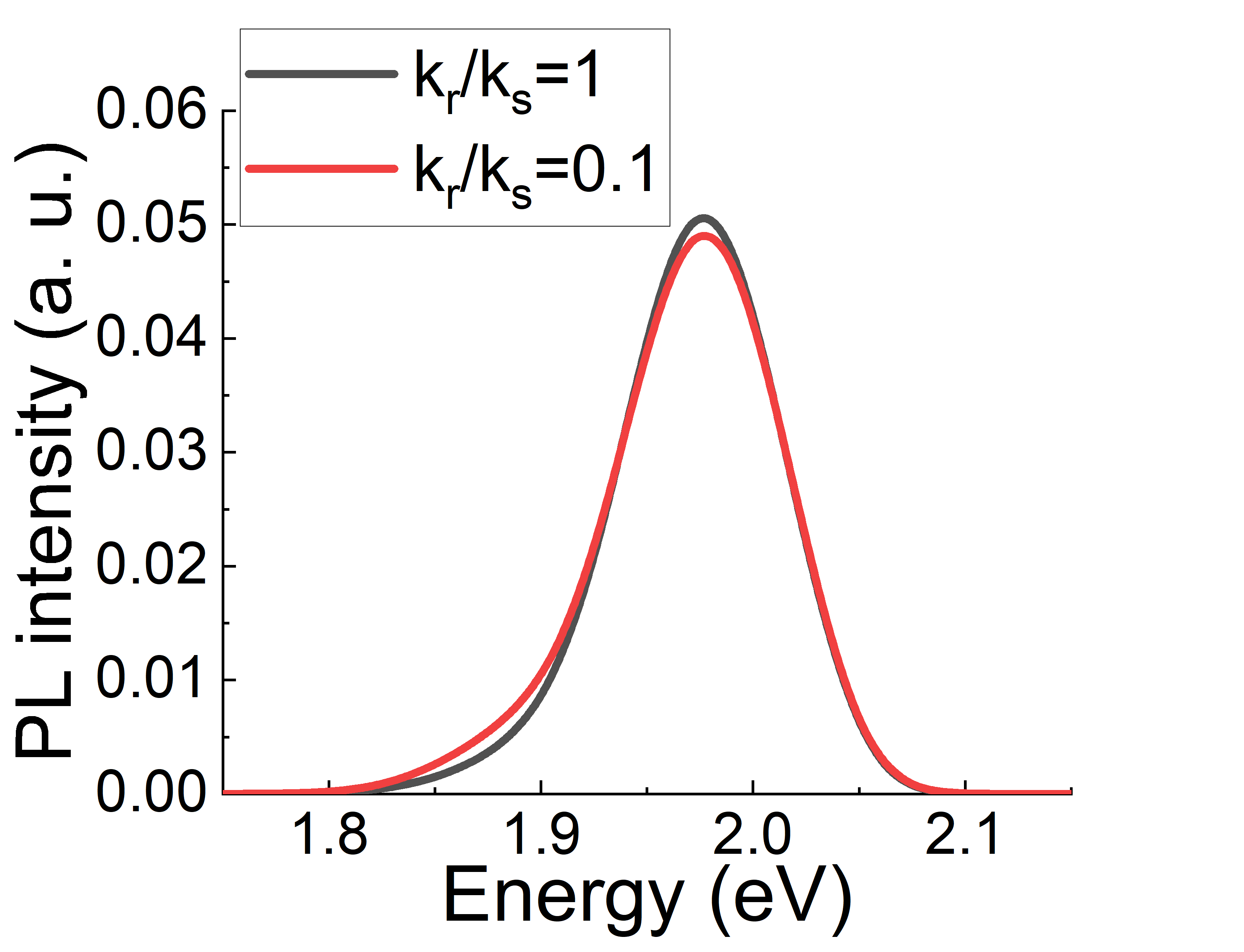}
	    \label{fig:ksc}}
	    \quad
	\subfloat[]{%
	    \includegraphics[scale=0.28]{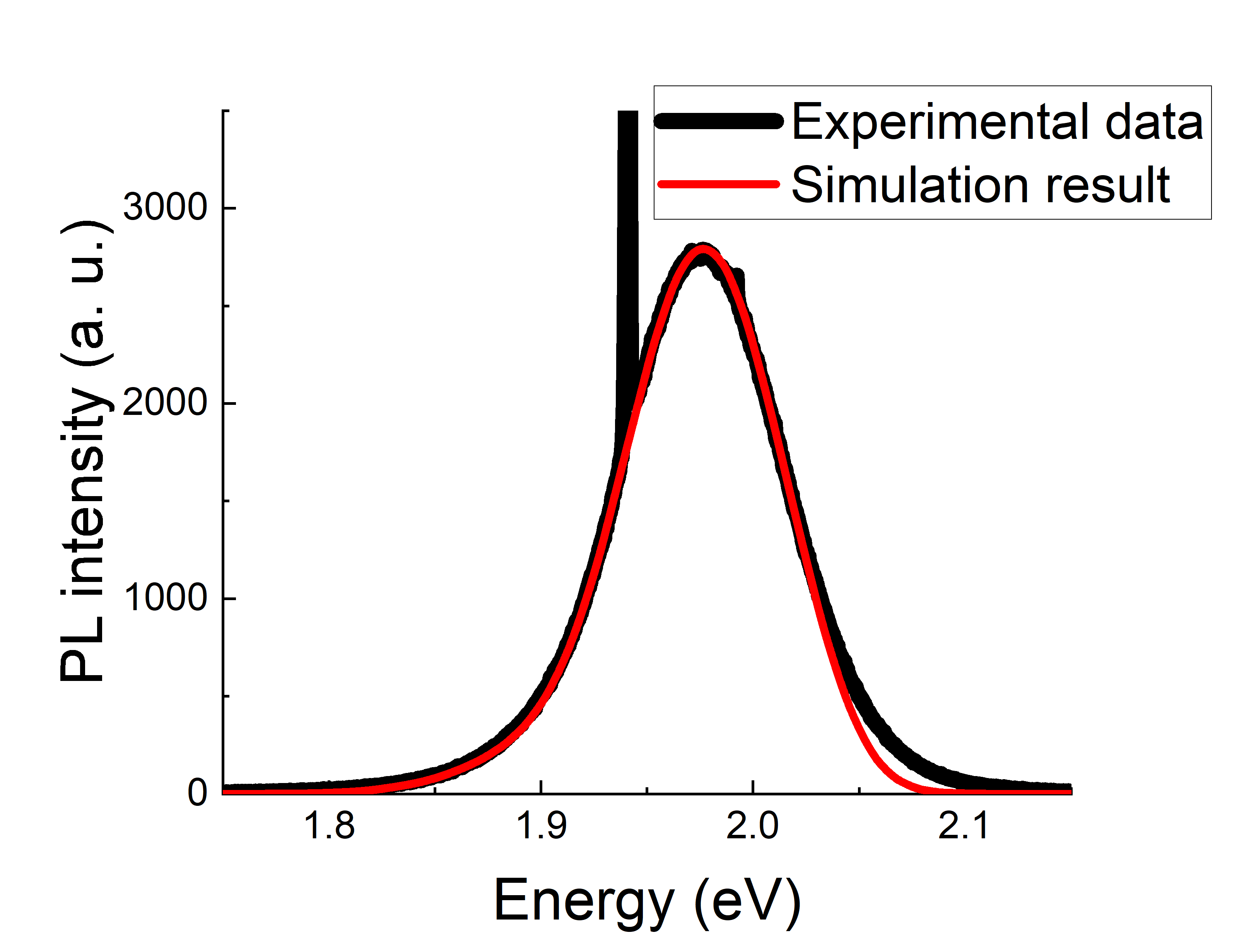}
	    \label{fig:fitting}}
	    \quad
	\subfloat[]{%
	    \includegraphics[scale=0.28]{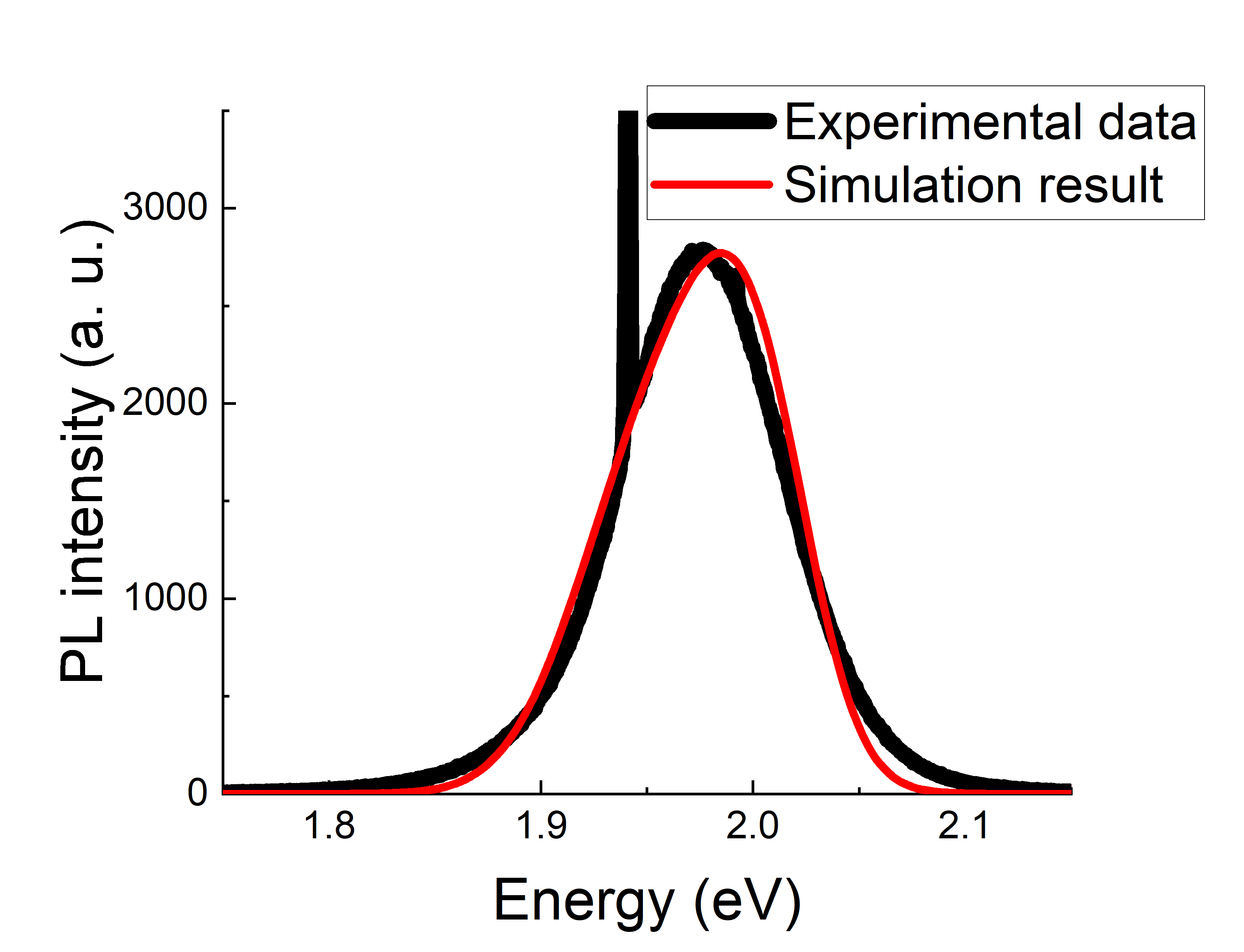}
	    \label{fig:n2fit}}
	    \quad
    \caption{Simulated PL spectra with (a) $k_{\textrm{r}}/k_{\textrm{s}}=1$ and (b) $k_{\textrm{r}}/k_{\textrm{s}}=0.5$, where an increased intensity indicates a larger iteration step. (c) Fully iterated PL spectra with $k_{\textrm{r}}/k_{\textrm{s}}=1$ and $k_{\textrm{r}}/k_{\textrm{s}}=0.1$. Simulation results fitted to the experimental data (sample 3 with $\varepsilon_{\textrm{ex}}=1.935\,\textrm{eV}$) with (d) $\textrm{n}=3$, $\varepsilon_{\textrm{s}}=1.976\,\textrm{eV}$ and $\sigma_{\textrm{s}}=38$~meV (FWHM=$89\,\textrm{meV}$), (e) $\textrm{n}=2$, $\varepsilon_{\textrm{s}}=2.000\,\textrm{eV}$ and $\sigma_{\textrm{s}}=34\,\textrm{meV}$ (FWHM=$80\,\textrm{meV}$).  In both simulations $\beta/k_{\textrm{r}}=0.01$ (ensuring the excitonic density is less than 1).}
\end{figure}

As shown in Fig. \ref{fig:ks1} and \ref{fig:ks05}, the recurrence converges as $\textrm{m}$ increases. Meanwhile, as expected, a slower converging speed was observed with smaller $k_{\textrm{r}}/k_{\textrm{s}}$ value. The simulation's $k_{\textrm{r}}/k_{\textrm{s}}$ dependence is shown in Fig. \ref{fig:ksc}, where UCPL(DCPL) processes were slightly reduced(enhanced) as $k_{\textrm{r}}/k_{\textrm{s}}$ decreases, indicating a longer radiative decay lifetime. As $k_{\textrm{r}}/k_{\textrm{s}}\rightarrow0$, it is invalid to treat $k_{\textrm{s}}O_{\textrm{s}}(\varepsilon)g_{\textrm{n}}(\varepsilon)\circledast N_{\textrm{s}}(\varepsilon)$ as a correction anymore. As $k_{\textrm{r}}/k_{\textrm{s}}$ decreases, the iteration itself could be understood as resembling the thermalization process. From the numerical calculation stand point, the computation time increases too fast as the number of iterations increases, consequently, our calculations stopped when $k_{\textrm{r}}/k_{\textrm{s}}=0.1$ .
The simulation result is shown in Fig. \ref{fig:fitting}, where the fitting was done by adjusting the joint width of the SET and VB bands, indicating a good resemble the experimental data. The relatively low intra-band transition rate can be understood as a poor bulk-phonon-mediated transition between the SET and CB states, compounded by the reduced mobility of the surface carrier due to the surface passivation treatment. 
\begin{figure}
    \subfloat[]{%
        \includegraphics[scale=0.28]{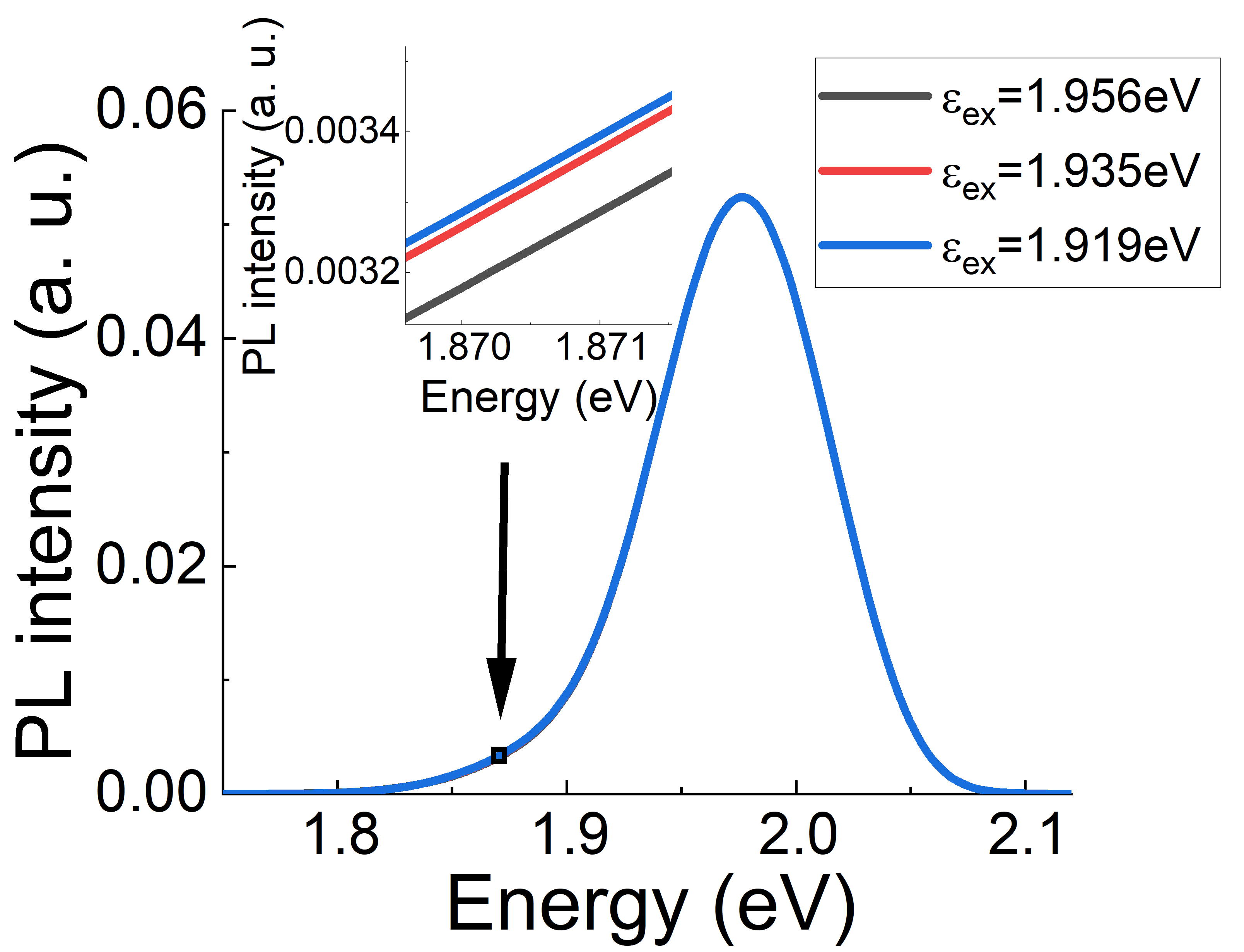}
        \label{fig:sexd}}%
        \quad
    \subfloat[]{%
	    \includegraphics[scale=0.28]{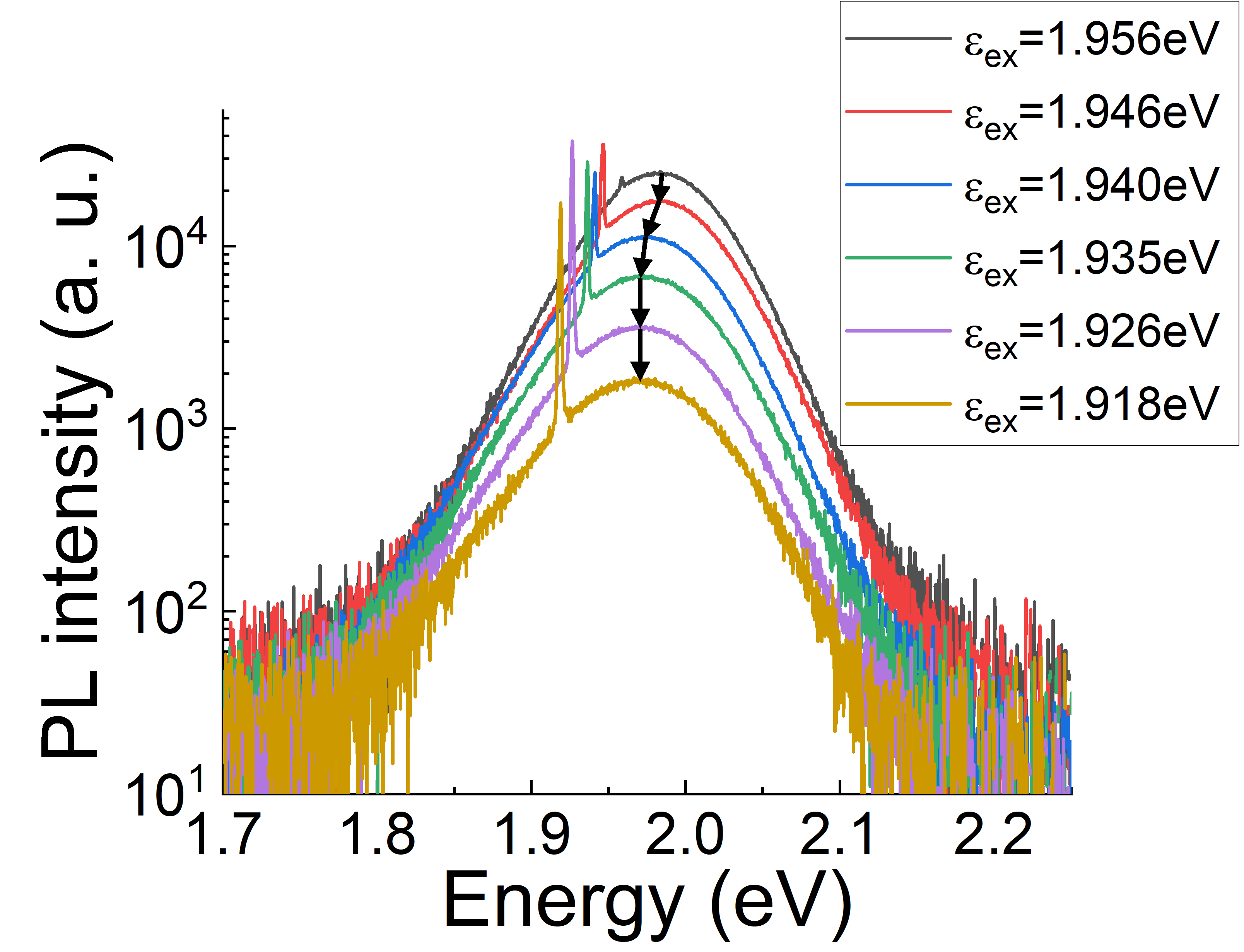}
	    \label{fig:ucplexd}}
	    \quad 
    \caption{(a) Simulated PL spectra ($9\,\textrm{th}$ iteration) with $\varepsilon_{\textrm{ex}}=1.956\,\textrm{eV}$ (black) and $1.919\,\textrm{eV}$ (red), the insert is a close view of the small rectangular area, (b) PL spectra of the sample with $\varepsilon_{\textrm{ex}}=1.956\,\textrm{eV}$ (black), $1.945\,\textrm{eV}$ (red), $1.938\,\textrm{eV}$ (blue), $1.929\,\textrm{eV}$ (green) and $1.919\,\textrm{eV}$ (purple).}
\end{figure}

When investigating the model's excitation energy dependence, the simulated PL line shape is very insensitive to the change in $\varepsilon_{\textrm{ex}}$ (Fig. \ref{fig:sexd}). Different from the experimental data, where slightly red-shift of the samples' PL spectra was observed as $\varepsilon_{\textrm{ex}}$ decreases, negligible change was suggested by the simulation results. Such difference can be understood as the QD ensemble's PL line shape was convoluted by their size distribution, leading to a red-shift of the UCPL peak with decreased $\varepsilon_{\textrm{ex}}$. This effect is suppressed when lowering the excitation energy, as fewer QDs (only the largest ones) of the QDs participate in the PL processes, leading to a PL spectral narrowing.\cite{Norris:1995} The experimental data are shown in Fig. \ref{fig:ucplexd} (the PL intensity was plotted in logarithmic scale as it decreases rapidly when $\varepsilon_{\textrm{ex}}$ decreases), the red-shift of the UCPL peak energy gradually stops as $\varepsilon_{\textrm{ex}}$ decreases. This is a strong indication that the model resembles the UCPL processes in QDs. One more fact that needs to be pointed out is that Fig. \ref{fig:ucplexd} also suggests that a deep trap emission is not observable even at SBE, further suggesting the lack of defects in the cores.

The case where the thermalization processes is mainly due to surface phonons was also tested, using $\textrm{n}=2$ in Eq. \ref{eq:phc},
\begin{equation}
    g_{2}(\varepsilon)=C_{\textrm{g}}\left\lvert\frac{\varepsilon}{e^{\varepsilon/k_{\textrm{B}}T}-1}\right\rvert.
    \label{eq:g2d}
\end{equation}
With the 2-D phonon mod, the pre-determined $\varepsilon_{\textrm{s}}$ and $\sigma_{\textrm{s}}$ values were not able to reproduce the experimental results. They have to be treated as fitting parameters and a higher $\varepsilon_{\textrm{s}}\approx2.000\,\textrm{eV}$ and a narrower $\sigma_{\textrm{s}}\approx34\,\textrm{meV}$ were required in the simulation, regardless of the value of $k_{\textrm{r}}/k_{\textrm{s}}$ value. Such observation suggests that if both 3-D and 2-D phonon mods were coupled in the QDs' PL processes at SBE, they will peak at different energies. When calculating $\chi^2$ (defined as $\chi^2=\Sigma\frac{(I_{\textrm{s}}(\varepsilon)-I_{\textrm{e}}(\varepsilon))^2}{I_{\textrm{e}}(\varepsilon)}$, where $I_{\textrm{s}}(\varepsilon)$ and $I_{\textrm{e}}(\varepsilon)$ denote the PL intensity at energy $\varepsilon$ obtained with the simulated result and experimental data respectively. The integrated PL intensities were normalized to unity before running the calculation), the 3-D model ($\chi^2\approx0.044$) was found to be slightly better than the 2-D model($\chi^2\approx0.062$). However, as shown in the Fig. \ref{fig:n2fit}, the fitting result produced with the 2-D model is noticeably worse than the 3-D ones. As a conclusion, although the existence of the surface phonon coupled processes cannot be completely excluded, our model suggests that they are not the major processes to produce UCPL in CdSe/CdS QDs.

\begin{figure}
    \centering
    \subfloat[]{%
	\includegraphics[scale=0.26]{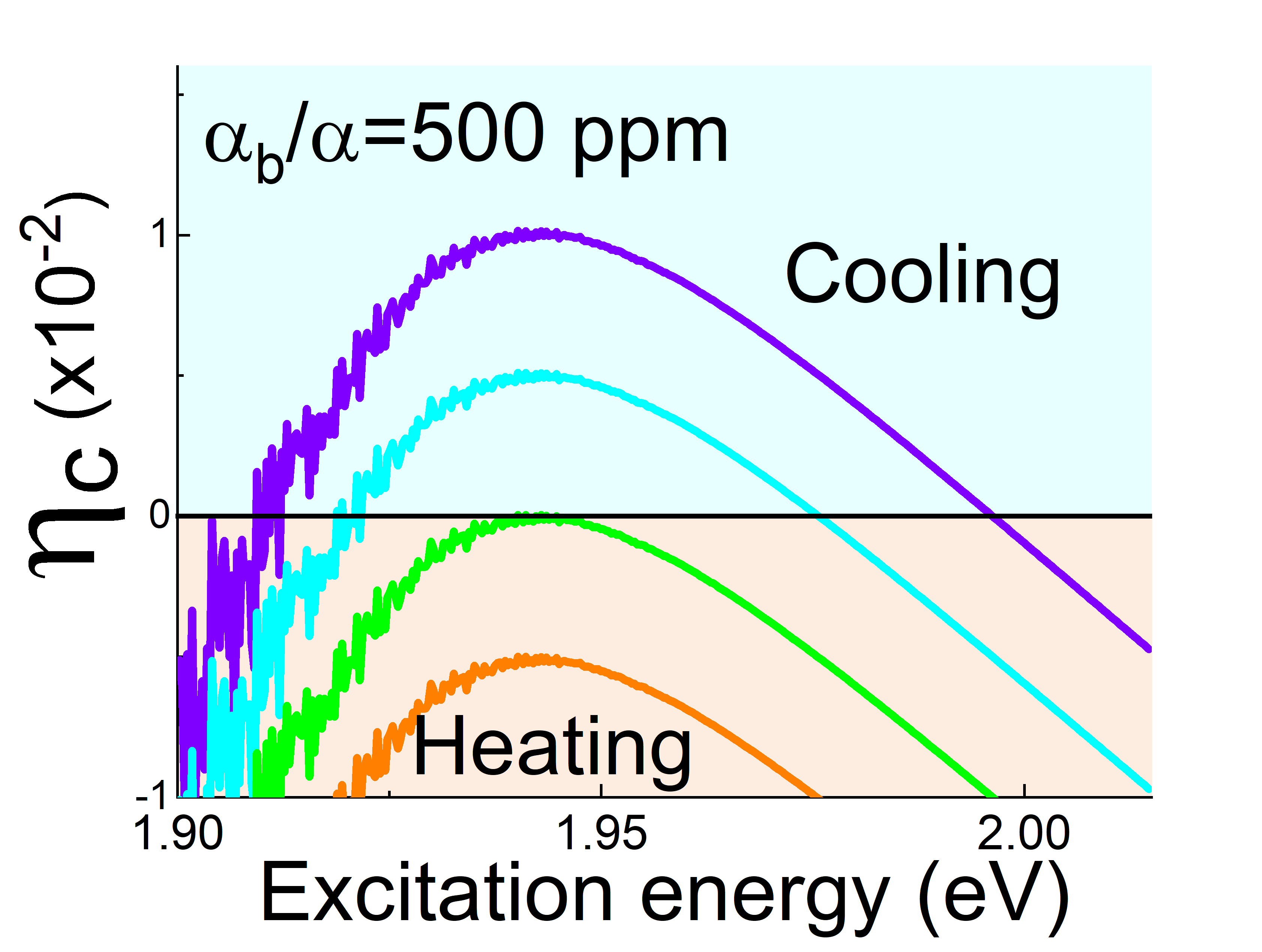}
	\label{ceff}}
    \quad
    \subfloat[]{%
	\includegraphics[scale=0.26]{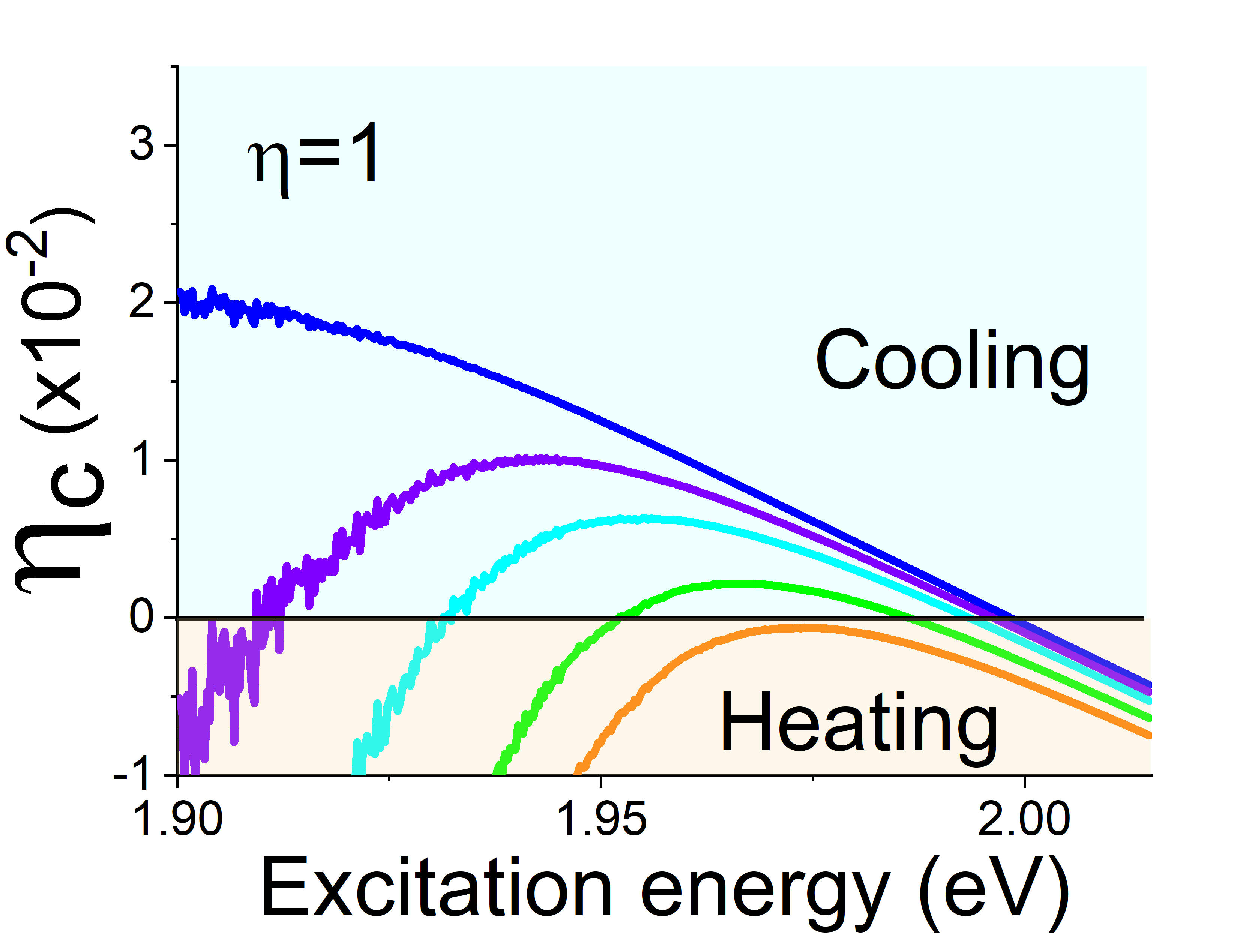}
	\label{ceffa}}
    \caption{Estimated cooling efficiency of sample $2$ as a function of $\varepsilon_{\textrm{ex}}$. (a) From top-to-bottom: $\eta=1$ , $0.995$ , $0.99$  and $0.985$. (b) From top-to-bottom:  $\frac{\alpha_{\textrm{b}}}{\alpha}=100$~ppm , $500$~ppm , $1000$~ppm , $2000$~ppm  and $3000$~ppm . $\alpha$ is obtained from the absorption spectrum.}
    \label{ce}
\end{figure}

In conclusion, except for the UCPL high energy tail, the model provides robust predictions of the PL line shape. Such disagreement could possibly due to the over simplification of the model that no exciton detrapping processes (electron detrapping from SET to conduction band) are considered. Since the difference of the ill fitted high energy tail is insignificant comparing to the overall line shape and would eventually lead to an underestimation of the net energy UC, the simulated PL intensity spectrum $I(\varepsilon)$ can be used to calculate the mean emission energy $\bar{\varepsilon}_{\textrm{em}}$ given by
\begin{equation}
    \bar{\varepsilon}_{\textrm{em}}=\frac{\int_{0}^{\infty}\varepsilon I(\varepsilon)d\varepsilon}{\int_{0}^{\infty}I(\varepsilon)d\varepsilon}.
    \label{mee}
\end{equation}
where $\int_{0}^{\infty}\varepsilon I(\varepsilon)d\varepsilon$ denotes the total emission power and $\int_{0}^{\infty}I(\varepsilon)d\varepsilon$ denotes the total emission rate. Plugging the result of Eq. \ref{mee} into Eq. \ref{eff}, $\eta_{c}$ can be obtained. Fig. \ref{ce} shows the results for sample $2$'s possible cooling efficiency with different QY $\eta$ and background absorption strength $\frac{\alpha_{\textrm{b}}}{\alpha}$. A cooling zone is predicted when $\eta>0.985$ and $\frac{\alpha_{\textrm{b}}}{\alpha}$ less than $2000$~ppm. The calculation also suggests that unless the background absorption is very small, a maximum of $\eta_{c}$ is found around $1.94$~eV. Since the synthesis method used for our QDs samples is supposed to yield almost unity QY, and it is supported by all experimental data, a net cooling effect might have already taken place during the experiment.

\section{Conclusions}\label{conclusions}
UCPL spectra of zinc-blended CdSe/CdS QDs were obtained with SBE. The experimental data suggested that the PL processes were mainly due to optical transitions between the SET and VBE, while the energy UC was achieved through the thermalization processes within dispersed SET. Therefore, a net energy UC is, in principle, achievable by exciting the QDs at the tail of their SET. Based on experimental data, a semi-empirical model describing the PL processes was constructed. The simulated results show extraordinary agreement with the experimental data and further indicated that the broadening of zinc-blende CdSe QDs' UCPL spectra is mainly due to the strong coupling between the photon induced exciton and the ``bulk" acoustic phonon modes. They also suggest that surface passivation reduces excitons' mobility in the surface band. Although the large cutoff phonon energy cannot yet be explained, quantitative predictions can be produced by the model. With the model's help, the possible cooling efficiency in CdSe QDs were calculated, showing the possibility of realizing OR in high quality zinc-blende CdSe QDs. 

\section{Acknowledgements}
We want to thank Dr. Daniel Minner, who provided generous practical suggestions in QDs synthesis. Dr. Bruce Ray, who helped us construct the synthesis equipment. We also want thank Professor Vemuri in our department, who kindly provided helpful advise in orienting the paper structure. R.~S.~D. acknowledges support from the National Science Foundation through grant PHY-1607360 and technical support from the IUPUI Integrated Nanosystem Development Institute.

\bibliographystyle{h-physrev}
\bibliography{Cooling_efficiency.bib}

\end{document}